\documentclass[twoside,twocolumn,9pt]{article}
\usepackage{extsizes}
\usepackage[super,sort&compress,comma]{natbib} 
\usepackage[version=3]{mhchem}
\usepackage[left=1.5cm, right=1.5cm, top=1.785cm, bottom=2.0cm]{geometry}
\usepackage{balance}
\usepackage{mathptmx}
\usepackage{sectsty}
\usepackage{graphicx} 
\usepackage{lastpage}
\usepackage[format=plain,justification=justified,singlelinecheck=false,font={stretch=1.125,small,sf},labelfont=bf,labelsep=space]{caption}
\usepackage{float}
\usepackage{fancyhdr}
\usepackage{fnpos}
\usepackage[english]{babel}
\addto{\captionsenglish}{%
  \renewcommand{\refname}{Notes and references}
}
\usepackage{array}
\usepackage{droidsans}
\usepackage{charter}%https://www.overleaf.com/project/61dffb6efbe622ac184c2748
\usepackage[T1]{fontenc}
\usepackage[usenames,dvipsnames]{xcolor}
\usepackage{setspace}
\usepackage[compact]{titlesec}
\usepackage{hyperref}
\usepackage{epstopdf}%This line makes .eps figures into .pdf - please comment out if not required.

\definecolor{cream}{RGB}{222,217,201}

\begin{document}

\pagestyle{fancy}
\thispagestyle{plain}
\fancypagestyle{plain}{
%%%HEADER%%%
\renewcommand{\headrulewidth}{0pt}
}
%%%END OF HEADER%%%

%%%PAGE SETUP - Please do not change any commands within this section%%%
\makeFNbottom
\makeatletter
\renewcommand\LARGE{\@setfontsize\LARGE{15pt}{17}}
\renewcommand\Large{\@setfontsize\Large{12pt}{14}}
\renewcommand\large{\@setfontsize\large{10pt}{12}}
\renewcommand\footnotesize{\@setfontsize\footnotesize{7pt}{10}}
\makeatother

\renewcommand{\thefootnote}{\fnsymbol{footnote}}
\renewcommand\footnoterule{\vspace*{1pt}% 
\color{cream}\hrule width 3.5in height 0.4pt \color{black}\vspace*{5pt}} 
\setcounter{secnumdepth}{5}

\makeatletter 
\renewcommand\@biblabel[1]{#1}            
\renewcommand\@makefntext[1]% 
{\noindent\makebox[0pt][r]{\@thefnmark\,}#1}
\makeatother 
\renewcommand{\figurename}{\small{Fig.}~}
\sectionfont{\sffamily\Large}
\subsectionfont{\normalsize}
\subsubsectionfont{\bf}
\setstretch{1.125} %In particular, please do not alter this line.
\setlength{\skip\footins}{0.8cm}
\setlength{\footnotesep}{0.25cm}
\setlength{\jot}{10pt}
\titlespacing*{\section}{0pt}{4pt}{4pt}
\titlespacing*{\subsection}{0pt}{15pt}{1pt}
%%%END OF PAGE SETUP%%%

%%%FOOTER%%%
\fancyfoot{}
\fancyfoot[LO,RE]{\vspace{-7.1pt}\includegraphics[height=9pt]{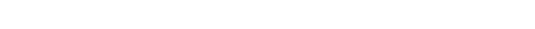}}
\fancyfoot[CO]{\vspace{-7.1pt}\hspace{13.2cm}\includegraphics{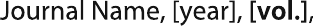}}
\fancyfoot[CE]{\vspace{-7.2pt}\hspace{-14.2cm}\includegraphics{head_foot/RF}}
\fancyfoot[RO]{\footnotesize{\sffamily{1--\pageref{LastPage} ~\textbar  \hspace{2pt}\thepage}}}
\fancyfoot[LE]{\footnotesize{\sffamily{\thepage~\textbar\hspace{3.45cm} 1--\pageref{LastPage}}}}
\fancyhead{}
\renewcommand{\headrulewidth}{0pt} 
\renewcommand{\footrulewidth}{0pt}
\setlength{\arrayrulewidth}{1pt}
\setlength{\columnsep}{6.5mm}
\setlength\bibsep{1pt}
%%%END OF FOOTER%%%

%%%FIGURE SETUP - please do not change any commands within this section%%%
\makeatletter 
\newlength{\figrulesep} 
\setlength{\figrulesep}{0.5\textfloatsep} 

\newcommand{\topfigrule}{\vspace*{-1pt}% 
\noindent{\color{cream}\rule[-\figrulesep]{\columnwidth}{1.5pt}} }

\newcommand{\botfigrule}{\vspace*{-2pt}% 
\noindent{\color{cream}\rule[\figrulesep]{\columnwidth}{1.5pt}} }

\newcommand{\dblfigrule}{\vspace*{-1pt}% 
\noindent{\color{cream}\rule[-\figrulesep]{\textwidth}{1.5pt}} }

\makeatother
%%%END OF FIGURE SETUP%%%

%%%TITLE, AUTHORS AND ABSTRACT%%%
\twocolumn[
  \begin{@twocolumnfalse}
{\includegraphics[height=30pt]{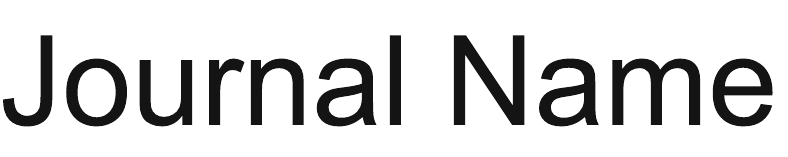}\hfill\raisebox{0pt}[0pt][0pt]{\includegraphics[height=55pt]{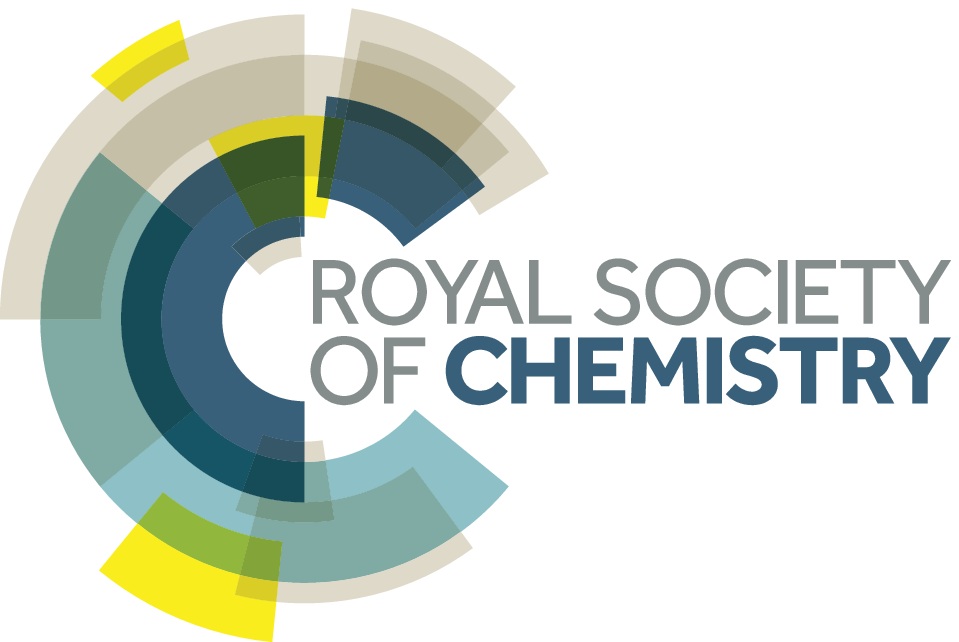}}\\[1ex]
\includegraphics[width=18.5cm]{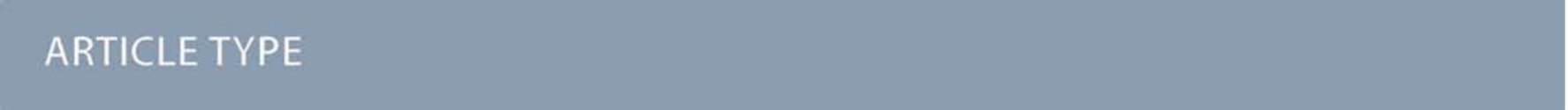}}\par
\vspace{1em}
\sffamily
\begin{tabular}{m{4.5cm} p{13.5cm} }

\includegraphics{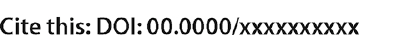} & \noindent\LARGE{\textbf{Nonadiabatic forward flux sampling for excited-state rare events$^\dag$
%Nonadiabatic forward flux sampling with surface hopping for excited-state rare events
}} \\%Article title goes here instead of the text ''This is the title''
\vspace{0.3cm} & \vspace{0.3cm} \\

 & \noindent\large{Madlen Maria Reiner,\textit{$^{ab}$} Brigitta Bachmair,\textit{$^{ac}$} Maximilian Xaver Tiefenbacher,\textit{$^{ac}$} Sebastian Mai,\textit{$^d$} Leticia González,$^\ast$\textit{$^{ad}$}} Philipp Marquetand,$^\ast$\textit{$^{ad}$} and  Christoph Dellago$^\ast$\textit{$^{ae}$} \\%Author names go here instead of ''Full name'', etc.

\includegraphics{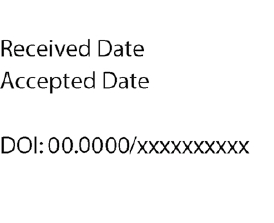} & \noindent\normalsize{We present a rare event sampling scheme applicable to coupled electronic excited states. In particular, we extend the forward flux sampling (FFS) method for rare event sampling to a nonadiabatic version (NAFFS) that uses the trajectory surface hopping (TSH) method for nonadiabatic dynamics. 
%Excited-state dynamics of molecules have been successfully studied using TSH before. However, such simulations carried out on highly accurate \emph{ab initio} potentials are computationally unaffordable for time scales beyond picoseconds. Hence, they do not allow the study of reactions that include rare barrier crossing events. 
%The FFS algorithm circumvents this issue in the ground state by directly sampling reactive paths and has been broadly applied to various problems in the literature. Hence, in the ground state the rare event sampling problem is conceptually solved, however, in exited-state systems it remains an issue. The presented combination of TSH and our newly developed excited-state FFS version allows the investigation of time scales bridging nonadiabatic reactions including rare events. 
NAFFS is applied to two dynamically relevant excited-state models that feature an avoided crossing and a conical intersection with tunable parameters. 
We investigate how nonadiabatic couplings, temperature, and reaction barriers affect transition rate constants in regimes that cannot be otherwise obtained with plain, traditional TSH. 
The comparison with reference brute-force TSH simulations for limiting cases of rareness shows that NAFFS can be several orders of magnitude cheaper than conventional TSH, and thus represents a conceptually novel tool to extend excited-state dynamics to time scales that are able to capture rare nonadiabatic events.

} 

\end{tabular}

 \end{@twocolumnfalse} \vspace{0.6cm}

  ]
%%%END OF TITLE, AUTHORS AND ABSTRACT%%%

%%%FONT SETUP - please do not change any commands within this section
\renewcommand*\rmdefault{bch}\normalfont\upshape
\rmfamily
\section*{}
\vspace{-1cm}

%%%FOOTNOTES%%%

\footnotetext{\textit{$^{a}$~Research Platform on Accelerating Photoreaction Discovery (ViRAPID), University of Vienna, Währinger Straße 17, 1090 Vienna, Austria. E-mail:
leticia.gonzalez@univie.ac.at,
philipp.marquetand@univie.ac.at, christoph.dellago@univie.ac.at}}
\footnotetext{\textit{$^{b}$~Vienna Doctoral School in Physics, University of Vienna, Boltzmanngasse 5, 1090 Vienna, Austria. }}
\footnotetext{\textit{$^{c}$~Vienna Doctoral School in Chemistry, University of Vienna, Währinger Straße 42, 1090 Vienna, Austria. }}
\footnotetext{\textit{$^{d}$~Institute of Theoretical Chemistry, Faculty of Chemistry, University of Vienna, Währinger Straße 17, 1090 Vienna, Austria. }}
\footnotetext{\textit{$^{e}$~Faculty of Physics, University of Vienna, Kolingasse 14-16, 1090 Vienna, Austria. }}

%Please use \dag to cite the ESI in the main text of the article.
%If you article does not have ESI please remove the the \dag symbol from the title and the footnotetext below.
\footnotetext{\dag~Electronic Supplementary Information (ESI) available: See DOI: 00.0000/00000000.}
%additional addresses can be cited as above using the lower-case letters, c, d, e... If all authors are from the same address, no letter is required

%\footnotetext{\ddag~Additional footnotes to the title and authors can be included \textit{e.g.}\ `Present address:' or `These authors contributed equally to this work' as above using the symbols: \ddag, \textsection, and \P. Please place the appropriate symbol next to the author's name and include a \texttt{\textbackslash footnotetext} entry in the the correct place in the list.}

%%%END OF FOOTNOTES%%%

%%%MAIN TEXT%%%%
\section{Introduction}

%% LG  to add and comment? doi/10.1098/rsta.2020.0382

Chemical reactions initiated by the absorption of a photon are at the core of organic synthesis,\cite{Honda2012} catalysis,\cite{Candish2022} optogenetics,\cite{Feliu2018}  protein modification,\cite{Holland2020} the conversion and storage of solar energy,\cite{Tian2018} and hold promise for many other applications.\cite{Ciamician1912}
While many photochemical reactions are ultrafast and occur on a femtosecond time scale,\cite{DeNalda2013} high barriers on electronically excited potential energy surfaces (PESs) or nonadiabatic transitions\cite{Domcke2011} with small couplings may lead to much slower reactions. For instance, the average reaction time for the keto-enol tautomerism of $2$-benzylbenzophenone is half a millisecond,\cite{Turro1965,Zwicker1963} \textit{i.e.}, many orders of magnitude longer than the time scale of basic molecular motions. The resulting separation of time scales represents a huge challenge for the computer simulation of such rare reactive events. In particular, the femtosecond time step\cite{Hammes-Schiffer1995} needed to accurately capture nuclear dynamics\cite{Dellago2009a,Lindh2020} makes it unfeasible to simulate rare photoreactions using straightforward quantum dynamics\cite{Curchod2018,Reiter2020} or nonadiabatic mixed quantum-classical molecular dynamics,\cite{Crespo-Otero2018,Mai2020b} even if recently developed machine learning approaches bring such simulations from the picosecond\cite{Zobel2021a} to the nanosecond time scale.\cite{Westermayr2019,Li2021,Westermayr2020MLST_Perspective,Westermayr2021CR}

For classical dynamics in the electronic ground state, numerous computational methods have been developed to address the rare event problem,\cite{Dellago2009a} including umbrella sampling,\cite{TORRIE1977187} blue-moon sampling,\cite{Ciccotti2004} steered MD,\cite{Tavan1996} hyperdynamics,\cite{Voter1997} milestoning,\cite{Elber2004} metadynamics,\cite{Laio2002} the string method, \cite{String2002} transition path sampling (TPS),\cite{Dellago1998a} and forward flux sampling (FFS).\cite{Allen2005,Allen2006,Allen2006a} The investigation of rare events in excited-state problems, however, is much less explored. Recent efforts relied on metadynamics to probe intersection crossing points between adiabatic electronic states that lead to the slow formation of photoproducts.\cite{Pieri2021,Aldaz2018,Lindner2019} In other work, TPS was used to sample the nonadiabatic dynamics of open quantum systems as described by a quantum master equation preserving detailed balance.\cite{Schile2018} TPS has also been applied to semiclassical pathways\cite{Sherman2016} obtained by trajectory surface hopping (TSH).\cite{Tully1990} In contrast to other rare event methods, TPS does not require any prior knowledge of the reaction mechanism in terms of a reaction coordinate. The backward propagation of trajectories required in TPS, however, is not possible in the framework of TSH.\cite{Subotnik2015} While this difficulty can be circumvented by generating reverse trajectories with approximate quantum weights and subsequently reweighting them,\cite{Sherman2016,Tully1995} this procedure reduces the efficiency of the TPS simulation.

In this paper, we show how rare but important events occurring in electronic excited states can be studied with FFS, a trajectory based approach originally developed for driven non-equilibrium stochastic systems with unknown stationary phase space distribution. In this approach, a sequence of non-intersecting interfaces between reactants and products is used to sample the ensemble of transition paths and calculate reaction rate constants. We combine the FFS methodology with TSH dynamics, exploiting that FFS requires only forward integration of the equations of motion. Hence, it is not affected by the lack of time reversal symmetry of TSH. As demonstrated using two simple illustrative models, the novel nonadiabatic FFS (NAFFS) method presented here provides a general approach for enhanced path sampling in electronic excited states and allows studying rare nonadiabatic reactions on time scales exceeding by far those accessible with brute-force TSH simulations.

The remainder of the paper is organized as follows. In Sec.~\ref{sec:theory} we lay out the FFS algorithm and describe how it is combined with TSH. Details on the implementation of the method are provided in Sec.~\ref{sec:implementation}. Results obtained for two simple model systems are presented and discussed in Sec.~\ref{sec:results} and conclusions are provided in Sec.~\ref{sec:conclusions}.

\section{Theory}
\label{sec:theory}

In the following, we review the main concepts behind the FFS method and the TSH algorithm and explain how they had to be extended in order to be combined into NAFFS. 
\subsection{Forward flux sampling of rare events in electronically excited states}
\label{sec:FFS}
Rare event sampling methods for ground state problems sample regions of the phase space that are unlikely to be visited by standard MD calculations.
%because of the large separation of time scales between the small fundamental time scale that is necessary for the MD simulation, (\textit{i.e.}, a time step of $\sim 1$~fs) and the low frequency of the event. 
Among them, TPS approaches sample fully dynamical trajectories, (\textit{i.e.}, trajectories which could occur in exactly the same way in brute-force MD simulations with the correct probability) with Monte Carlo methods acting in path space (\textit{i.e.}, trajectory space).\cite{Dellago2009a} In these algorithms, transition paths---\textit{i.e.}, rare trajectories that start in the initial reactant region of phase space and end in the final product region of phase space---are sampled by generating a new trajectory from a given trajectory, typically by propagating the system both forward and backward in time. The newly generated trajectory is then accepted or rejected according to a criterion guaranteeing that the trajectories follow the statistics dictated by the transition path ensemble. In this way, reactivity is maintained at all times during the TPS simulation and no computing resources are wasted to follow the dynamics of the system during the long periods when no transition occurs. Analysis of the sampled transition paths then provides insights into the underlying reaction mechanisms.\cite{Dellago1998a} In transition interface sampling,\cite{VanErp2003} a variant of TPS, ensembles of pathways that cross a sequence of interfaces between reactant and product regions are considered. This procedure significantly enhances the sampling of transition paths and uses the rejected trajectories ingeniously to calculate reaction rate constants. 

As most other TPS methods, also transition interface sampling relies on the time reversibility of the dynamics and on explicit knowledge of the stationary phase space distribution. Hence, the application of TPS methods to irreversible non-equilibrium systems is not straightforward. This difficulty motivated the development of FFS,\cite{Allen2005,Allen2006,Allen2006a,Allen2009a} a simulation method for rare events in which the equations of motion are integrated only forward in time such that it does not suffer from lack of knowledge of the stationary distribution and the absence of microscopic reversibility. Hence, FFS, a splitting method based in spawning swarms of trajectories connecting interfaces, can be applied to study rare events occurring in non-equilibrium systems.\cite{Allen2009a,Bolhuis2021} These properties make FFS ideally suited for combining it with TSH, which does not satisfy detailed balance and generally produces an unknown stationary distribution. 

%when using the trajectory surface hopping approach for excited-state MD simulations, as meaningful backpropagation using TSH remains an unsolved issue for long simulations that include several hopping events.\cite{Subotnik2015} These problems motivate the choice of FFS as rare event path sampling algorithm for our methodology. We extend this method to excited-state systems in order to study nonadiabatic rare reaction events.

\begin{figure}
    \centering
    \includegraphics[height=4.7cm]{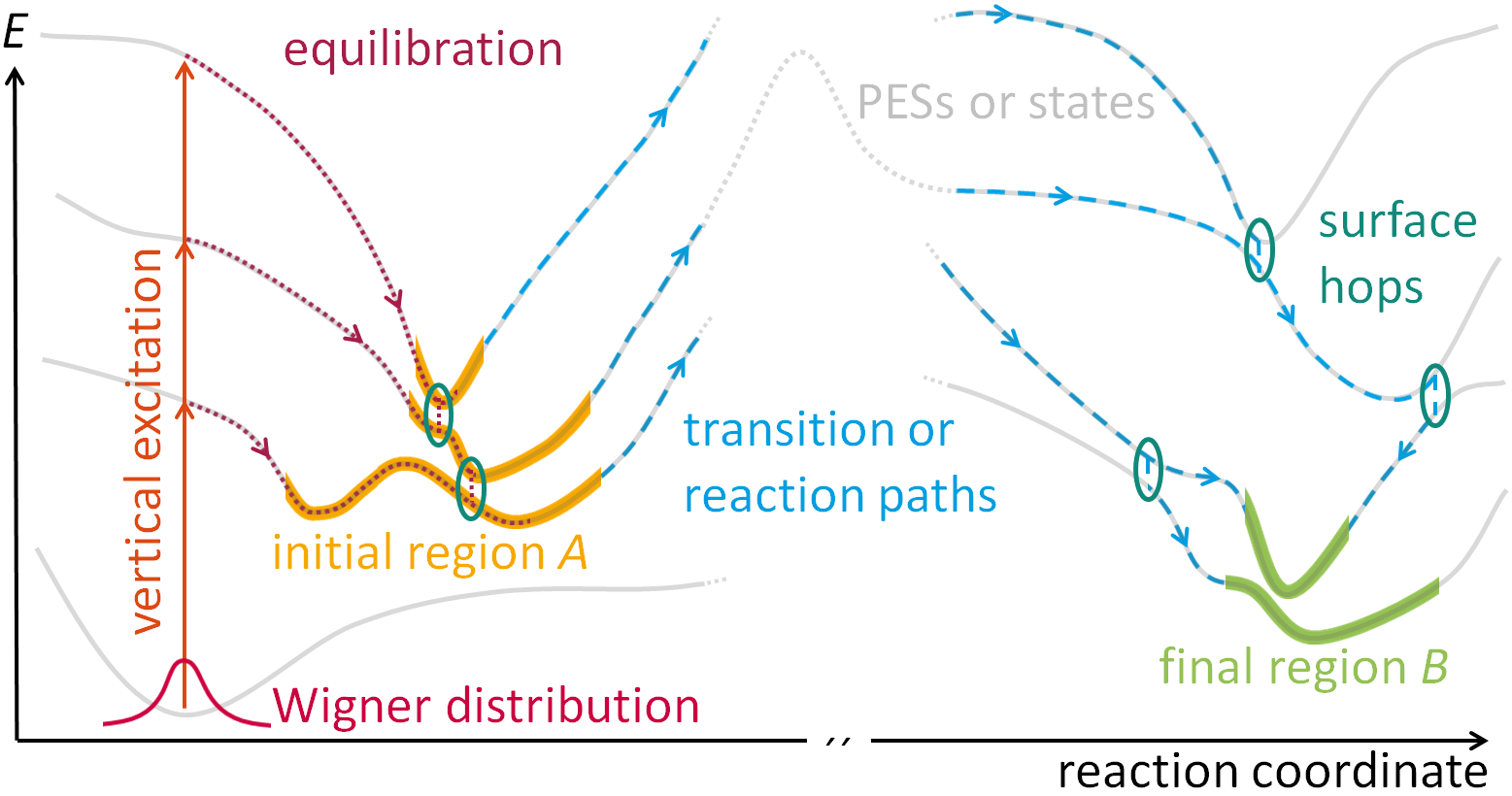}
    \caption{Schematic illustration of a nonadiabatic PES landscape, showing a possible definition of the initial and final regions $A$ and $B$. In the framework of a typical TSH simulation, a nuclear ensemble of vertically excited configurations drawn from the Wigner distribution relaxes into one or several minima included in $A$ (equilibration), which serves as starting region for the NAFFS simulation. NAFFS then samples reaction paths that connect $A$ and $B$ including rare events in electronic excited states (\textit{e.g.}, an energy barrier).
    %, and is not shown here ($\prime \prime$, see Fig.~\ref{fgr:FFS}). 
    Due to the nonadiabatic nature of the PES landscape, surface hops (circles) between states can happen during the dynamics.}
    \label{fgr:regions}
\end{figure}

As it is customary in FFS, we consider systems that undergo a rare transition from an initial reactant region of phase space called $A$ to a final product region called $B$. Both regions, $A$ and $B$, are supposed to be stable, meaning that the system resides in them for long times compared to the time where it is in an unstable region, \textit{e.g.}, when it undergoes a transition from $A$ to $B$.\cite{Dellago2009a} 
Both regions are defined in terms of collective variables, that is, functions of the phase space coordinates, \textit{e.g.}, bond lengths, angles, or dihedrals.\cite{Peters2016} 
What distinguishes our application of FFS from others is the novelty that we explicitly include one or several PESs in the definition of our initial and final regions (see Fig.~\ref{fgr:regions}); this is necessary to study nonadiabatic excited-state reactions.
Throughout this work, we use the term ``region'' for stable phase space regions in the FFS context, and the term ``transition path'' as a synonym of ``reaction path'' to describe the evolution of the system between regions.
We use the term ``state'' for electronic states or PESs, and the term ``hop'' to describe nonadiabatic changes between states.
Hence, in our nonadiabatic setting, one or several states can be part of the definition of a region, and a transition path between regions can include hops between electronic states.

A typical TSH simulation begins with an instantaneous vertical excitation (that mimics the absorption of a photon) of a nuclear ensemble of geometrical configurations, \textit{e.g.}, drawn from a Wigner distribution in the electronic ground state minimum (see Fig.~\ref{fgr:regions} left).\cite{Mai2020b} 
In excited-state reactions involving rare events, after the vertical excitation the system would evolve (see dark red dotted equilibration trajectories in Fig.~\ref{fgr:regions}) into a stable excited-state region $A$, in which it stays for a very long time before the rare event occurs and the system transitions to the final region $B$ (see blue dashed transition paths in Fig.~\ref{fgr:regions}).
To ensure the greatest possible generality in the choice of reaction pathways, the stable excited-state region $A$ may span several PESs, where the different configurations of the nuclear ensemble land after equilibration.
Further, the initial region $A$ is flexible enough to include different regions of a single PES, if needed.
Likewise, the final region $B$ (see green PESs parts in Fig.~\ref{fgr:regions}) can expand over multiple states.
For simplicity, neither the vertical excitation process nor the equilibration to the initial region $A$ is included in our NAFFS algorithm but can be performed with standard initial condition and TSH simulations.

\begin{figure*}
    \centering
    \includegraphics[height=2.75cm]{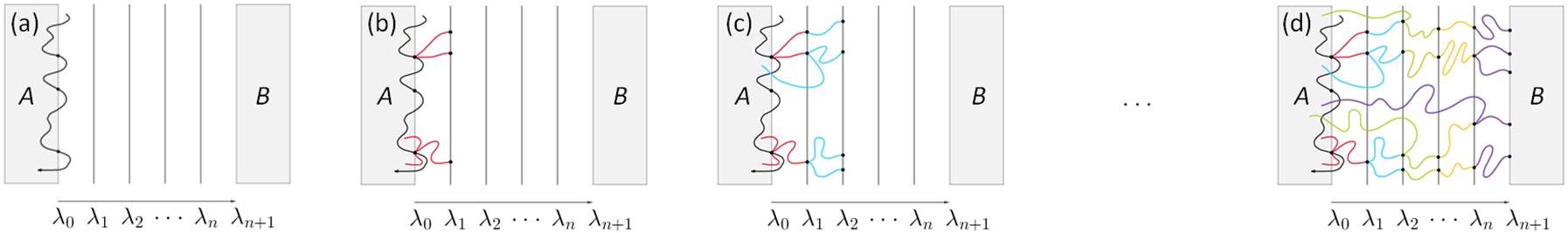}
    \caption{The working principle of the forward flux sampling scheme.\cite{Allen2009a} (a) A brute-force MD simulation in the initial reactant region $A$ is performed and the snapshots where the trajectory exits $A$ %in the direction of the final product region $B$ 
    are stored as initial shooting points. (b) Shots of randomly chosen initial shooting points are either rejected if they enter $A$, or accepted if they cross the adjacent interface $\lambda_1$. Due to the underlying stochastic dynamics, trajectories initiated in the same shooting point differ. Crossing snapshots are stored as shooting points for the next FFS cycle. (c) In the second FFS cycle, randomly chosen trajectories ending in shooting points on the interface $\lambda_1$ are continued, and either accepted if they cross the next adjacent interface, or rejected if they enter $A$. This procedure is repeated for all interfaces $\lambda_i$. (d) Once the last FFS cycle is finished, final transition paths from $A$ to $B$ are obtained by piecing together the accepted partial paths obtained in each FFS cycle.}
    \label{fgr:FFS}
\end{figure*}

As the transition interface sampling method,\cite{VanErp2003} FFS is an interface-based approach\cite{Escobedo2009} where interfaces $\lambda_i$ (\textit{i.e.}, intermediate stages between the regions $A$ and $B$) are defined in terms of collective variables.
There exist several approaches to the proper placement of the interfaces.\cite{Berkov2021,Borrero2011,Hussain2020,Bolhuis2015}
In NAFFS, the interfaces are also able to include a range of PESs, as the initial and final regions do.
For the applications presented later, we define the first and last interfaces to equal the boundaries of the stable regions $A$ and $B$, \textit{i.e.}, $\lambda_A \equiv \lambda_0$ and $\lambda_B \equiv \lambda_{n+1}$, as often done so.\cite{Allen2006a}
The rate constant $k_{AB}$ is then calculated as\cite{Allen2005}
\begin{equation}
    k_{AB}=\phi_{A}P_{A}(\lambda_{n+1}|\lambda_{0})=\phi_{A}\prod\limits_{i=0}^{n}P_{A}(\lambda_{i+1}|\lambda_{i}),
    \label{eq:FFS-rate}
\end{equation}
where $\phi_{A}$ is the effective positive flux out of $A$ through the boundary of $A$, and $P_A(\lambda_j|\lambda_i)$ is the probability of a trajectory that started in $A$ crossing the interface $\lambda_j$ in the direction of $B$ given that it has already crossed the interface $\lambda_i$.
There are two common ways to calculate the flux $\phi_A$ through the boundary of the initial region $A$.
The first\cite{Hussain2020} involves running an MD simulation in region $A$ and counting the number of times $N_0$ that the interface $\lambda_0$ is crossed in the outward direction of $A$ divided by the simulation time (see Fig.~\ref{fgr:FFS}a).
Here it is assumed that the trajectory does not enter the final product region $B$ during the MD simulation, or if it does, it is immediately reset to $A$ and re-equilibrated.\cite{Allen2006} 
Alternatively, only for equilibrium systems and reversible reactions and if the trajectory visits both regions $A$ and $B$ several times in the MD run, the flux $\phi_A$ can be calculated by dividing the number $N_0$ by the time $T_{\mathcal{A}}$ the system has spent in the overall region $\mathcal{A}$ during the simulation,\cite{VanErp2003}
\begin{equation}
    \phi_A = \frac{N_0}{T_\mathcal{A}}.
    \label{eq:FFS_flux}
\end{equation}
The time $T_\mathcal{A}$ spent in the overall region $\mathcal{A}$ is not only the time the trajectory is located in region $A$, but also the time spent outside $A$ as long as $B$ is not reached.
If the trajectory enters $B$, $T_\mathcal{A}$ resumes counting after the trajectory exits $B$ and re-enters $A$.
Provided that both approaches to calculate the flux $\phi_A$ are applicable, one or the other could be computationally more efficient and which one to take is decided depending on the system to study.

The crossing probabilities $P_A(\lambda_{i+1}|\lambda_i)$ are given by the fraction of accepted Monte Carlo moves or ``shots'' that are initiated on the shooting interface $\lambda_i$ and cross the subsequent interface $\lambda_{i+1}$ with respect to the total number $M$ of trial shots initiated from $\lambda_i$. 
In the first FFS cycle, the shooting points on the boundary $\lambda_0$ of the initial region $A$ are randomly chosen from the $N_0$ crossing points collected in the flux simulation, exploiting the fact that due to the underlying stochastic dynamics of the system (induced typically by a thermostat), two shots beginning in the same point produce different trajectories (see Fig.~\ref{fgr:FFS}b). Final points of accepted shots in each FFS cycle serve as possible shooting points for the next FFS cycle (see Fig.~\ref{fgr:FFS}c).\cite{Allen2006}
Final reactive paths (see Fig.~\ref{fgr:FFS}d) are obtained in accordance to their correct weight in the transition path ensemble, \textit{i.e.}, the relative probabilities of transition paths with respect to the considered system, and, hence, those of reactive paths obtained in MD simulations are conserved.\cite{Dellago2006}
In summary, the difficult problem of finding a  reaction coordinate and a phase space probability distribution is replaced by the (usually) simpler task of defining reactant and product regions $A$ and $B$, and interfaces in between. 

%MD simulations in the respective regions can be informative and in the excited-state context this information can be useful to look at which region the system equilibrates to after excitation.

The relative error of the rate constant $k_{AB}$ obtained in an FFS simulation can be estimated  as\cite{Allen2006a,Hussain2020,Borrero2008}
\begin{equation}
\frac{\Delta k_{AB}}{k_{AB}}=\sqrt{\frac{1}{N_0}+\sum\limits_{i=0}^{n}\frac{1-P_A(\lambda_{i+1}|\lambda_i)}{P_A(\lambda_{i+1}|\lambda_i)\cdot M_i}},
\label{eq:FFSerror}
\end{equation}
where $\Delta k_{AB}$ is the standard deviation of the calculated rate constant $k_{AB}$, and $M_i$ is the number of shots performed starting from shooting points on interface $\lambda_i$. The relative error estimation given by Eq.~(\ref{eq:FFSerror}) corresponds to a Gaussian error propagation, taking into consideration an estimate of the relative error of the flux $\phi_A$ as
\begin{equation}
\frac{\Delta \phi_{A}}{\phi_{A}}=\frac{\sqrt{N_0}}{N_0},
\end{equation}
and the estimated errors of the crossing probabilities obtained for each shooting interface as
\begin{equation}
\Delta P_A(\lambda_{i+1}|\lambda_i)=\sqrt{\frac{P_A(\lambda_{i+1}|\lambda_i)\cdot (1-P_A(\lambda_{i+1}|\lambda_i))}{M_i}}.
\end{equation}

%In conclusion, the FFS method satisfies all the conditions that arise when we want to sample nonadiabatic rare events in an accelerated fashion using the TSH algorithm, such as not requiring backpropagation or no need to know the stationary distribution. Hence, this overcomes the issues of other approaches on nonadiabatic transition path sampling.\cite{Schile2018,Sherman2016} 
%%%%%%%%%%%%%%%%%%%%%%%%%%%%%%%%%%%%
\subsection{Stochastic excited-state molecular dynamics simulations using trajectory surface hopping}
\label{sec:TSH}

The application of the FFS method requires an MD algorithm to propagate the system in time. 
%For our approach this propagation must be able to cope with excited states in order to describe nonadiabatic dynamics, \textit{e.g.}, processes like internal conversion when states of the same spin multiplicity interact via nonadiabatic couplings or intersystem crossing when states of different spin multiplicity interact via spin-orbit couplings.\cite{Mai2018,Marian2012} 
Here, we use a velocity Verlet-type\cite{Verlet1967,Verlet1968,Gronbech-Jensen2014} algorithm with Langevin dynamics in combination with the surface hopping including arbitrary couplings (SHARC) approach.\cite{Richter2011,Mai2015,Mai2018} 
SHARC is an extension of Tully's fewest switches\cite{Tully1971} TSH method, able to describe on the same footing internal conversion between states of the same spin multiplicity via nonadiabatic couplings and intersystem crossing between states of different spin multiplicity via spin-orbit couplings. 

As a TSH method, SHARC is a mixed quantum-classical simulation technique, where the nuclei are considered classical particles and nonadiabatic effects are accounted for by including multiple PESs.\cite{Nelson2020}
Nuclei always follow the force corresponding to one single PES (the ``active state''), and instantaneous hops between adiabatic PESs mimic nonadiabatic changes, according to hopping probabilities based on the quantum mechanical evolution of the electronic populations of the different states.\cite{Mai2018} 
%Trajectory surface hopping is a mixed quantum-classical simulation technique that accounts for quantum mechanical effects in MD simulations by including excited adiabatic PESs in the calculations and allows transitions or ``hops'' between them.\cite{Nelson2020} 
%While the nuclei are considered classically, \textit{i.e.}, localized on a single PES, hops between adiabatic states can happen according to hopping probabilities that are based on the quantum mechanical evolution of the electronic populations of the different states.\cite{Mai2018} 
As the TSH algorithm treats the electrons quantum mechanically, it solves the electronic time-dependent Schrödinger equation\cite{Mai2015}
\begin{equation}
    i\hbar \frac{\mathrm{d}}{\mathrm{d}t}|\Psi \rangle = \hat{H}|\Psi\rangle,
    \label{eq:SE}
\end{equation}
where $\hat{H}$ is the electronic Hamilton operator, $\hbar$ the reduced Planck constant, and $|\Psi\rangle$ the electronic wave function, which in a linear combination of basis states $\psi_\alpha$ is written in terms of the coefficients $c_\alpha$,
\begin{equation}
    |\Psi \rangle =\sum\limits_\alpha c_\alpha |\psi_\alpha \rangle.
    \label{eq:psi}
\end{equation}
Combining  Eq.~(\ref{eq:SE}) and Eq.~(\ref{eq:psi}) yields the equations of motion for the electronic population vector $\mathbf{c}$ consisting of the electronic wave function coefficients $c_\alpha$,
\begin{equation}
    \frac{\mathrm{d}}{\mathrm{d}t} \mathbf{c} = -\left[\frac{i}{\hbar} H + K \right]\mathbf{c}
\end{equation}
with coupling matrix $K$ and Hamilton matrix $H$.

SHARC uses a fully ``adiabatic'' or diagonal representation in the propagation.\cite{Richter2011,Mai2015} 
%We discuss crucial aspects of the resulting dynamics regarding our work in the following, where we consider everything in the diagonal,\cite{Richter2011} \textit{i.e.}, ``adiabatic'', representation unless stated differently. 
In the simulations presented in this work, the coefficients $c_\alpha$ are propagated as\cite{Mai2018}
\begin{equation}
    \mathbf{c}(t+\Delta t)=R(t+\Delta t,t)\cdot \mathbf{c}(t)
\end{equation}
with time step $\Delta t$.
The corresponding propagator matrix $R$ is given as
\begin{equation}
    R(t+\Delta t,t)=S(t,t+\Delta t)^\dagger \prod\limits_{i=1}^{s}R_i,
    \label{eq:propagator}
\end{equation}
with
\begin{equation}
    R_i = \exp\left[ -\frac{i}{\hbar} H_i \Delta \tau \right],
\end{equation}
and
\begin{equation}
    H_i = H(t) +\frac{i}{s}\left( S(t,t+\Delta t)H(t+\Delta t)S(t,t+\Delta t)^\dagger -H(t) \right),
\end{equation}
where $s$ the number of substeps with length $\Delta \tau = \frac{\Delta t}{s}$ within a time step of length $\Delta t$, 
%Planck's constant $\hbar$,  %% defined before
and overlap matrix $S$ calculated by $S(t,t+\Delta t)=U^\dagger (t)\cdot U(t+\Delta t)$ from transformation matrices $U$ obtained by diagonalizing the diabatic Hamiltonian, $H=U^\dagger H^{\textnormal{diab}}U$ at times $t$ and $t+\Delta t$. 
The overlap matrix and Hamiltonian are phase-corrected\cite{Mai2018} in each time step to avoid random changes in the population transfer direction because of random phases of the transformation matrices $U$. The hopping probabilities, \textit{i.e.}, the probabilities to hop from the current or active state $\beta$ to a different state $\alpha$, are given by\cite{Mai2015}
\begin{equation}
    P_{\beta \rightarrow \alpha}=\left( 1-\frac{|c_\beta (t+\Delta t)|^2}{|c_\beta (t)|^2} \right) \frac{\Re \left[ c_\alpha(t+\Delta t) R_{\alpha\beta}^\ast \left( c_\beta(t)\right)^\ast\right]}{|c_\beta (t)|^2-\Re \left[ c_\beta(t+\Delta t) R_{\beta\beta}^\ast \left( c_\beta(t)\right)^\ast \right]},
\end{equation}
with electronic coefficients in state $\alpha$ and $\beta$, namely $c_\alpha$ and $c_\beta$, and complex conjugated elements of the propagator matrix $R$. A surface hop to state $\widetilde{\alpha}$ is attempted only if a random number $r$ drawn from a uniform distribution in the interval $[0,1]$ satisfies\cite{Fabiano2008}
\begin{equation}
    \sum\limits_{i=1}^{\widetilde{\alpha} -1} P_{\beta\rightarrow i} < r \leq \sum\limits_{i=1}^{\widetilde{\alpha}} P_{\beta\rightarrow i}.
\end{equation}
If the total energy of the system is smaller than the potential energy of the envisaged new state $\widetilde{\alpha}$, no hop is performed and the system stays in the state $\beta$, \textit{i.e.}, the new active state $\beta$ is the same as the old state---this is called a ``frustrated hop''.
Otherwise, if the potential energy $E_{\widetilde{\alpha}}$ of the envisaged new state $\widetilde{\alpha}$ is smaller than or equal to the system's total energy $E_\textnormal{total}$, a surface hop $\beta \rightarrow \widetilde{\alpha}$ is performed.
As in most TSH implementations, during the surface hop the total energy is kept constant by rescaling 
%the kinetic energy $E_\textnormal{kin}$ of the system and 
the nuclear velocities $\mathbf{v}$. %is obtained by rescaling the entire velocity $\mathbf{v}$
By default in SHARC, this is done according to the scheme\cite{Mai2020b}
\begin{equation}
        \mathbf{v}^\prime = \sqrt{\frac{E_\textnormal{total}-E_{\widetilde{\alpha}}}{E_\textnormal{kin}}}\cdot \mathbf{v}, %\quad
%        E_\textnormal{kin}^\prime = \frac{E_\textnormal{total}-E_{\widetilde{\alpha}}}{E_\textnormal{kin}} \cdot E_\textnormal{kin},
    \label{eq:velrescaling}
\end{equation}
where $E_\textnormal{kin}$ is the total kinetic energy before the hop, \textit{i.e.}, the rescaled velocity vector $\mathbf{v}^\prime$ is parallel to the original one, $\mathbf{v}$.
After this adjustment, the state $\widetilde{\alpha}$ is the new active state $\beta$ of the system. Other velocity adjustment varieties are available in SHARC.\cite{Plasser2019a}

In TSH, electronic populations in the non-active states $\alpha$ follow the forces of the active state $\beta$, even though in a proper quantum mechanical description they should follow the forces of their respective state $\alpha$.
This problem is known as ``overcoherence'',\cite{Heindl2021,Crespo-Otero2018,Mai2020b} and in the present work is accounted for by modifying the electronic coefficients of the states $\alpha$ according to the energy difference to the active state $\beta$ after the surface hopping procedure,\cite{Granucci2010}
\begin{equation}
    \begin{split}
        c_{\alpha}^\prime & = c_\alpha \cdot \exp \left[ -\frac{1}{2}\Delta t \frac{|E_\alpha -E_\beta|}{\hbar} \left( 1+\frac{C}{E_\textnormal{kin}} \right)^{-1} \right], \quad \alpha\neq \beta,\\
        c_\beta^\prime & = \frac{c_\beta}{|c_\beta|} \left( 1-\sum\limits_{\alpha \neq \beta} |c_{\alpha^\prime}|^2 \right)^{-\frac{1}{2}}
    \end{split}
    \label{eq:decoherence}
\end{equation}
with decoherence parameter $C$.

Although TSH algorithms already have an intrinsically  stochastic character due to the randomness of the hops, their degree of stochasticity is not sufficient for an application of the FFS algorithm, especially in regions away from probable hopping points. 
These regions are characterized by large energy gaps between adjacent PESs. 
To increase the level of stochasticity beyond random hops in a controllable way, we consider a system evolving under the influence of friction and random forces as described by the Langevin equation
%To avoid problems in this regard, instead of propagating the nuclei based on the deterministic Newton's equations of motion, we use a Langevin thermostat that provides stochastic dynamics satisfying Langevin's equation
%
\begin{equation}
    m\dot{\mathbf{v}}=-\gamma \mathbf{v}(t) -\boldsymbol\nabla E(\mathbf{x}(t))+\mathbf{\eta}(t).
    \label{eq:Langevin}
\end{equation}
Here, $\mathbf{x}$ denotes the positions of all atoms, $E$ is the potential energy, $m$ the mass and $\gamma$ the friction constant controlling the magnitude of the frictional forces, which are proportional to the velocities. In the above equation, $\mathbf{\eta}(t)$ denotes Gaussian white noise with zero mean, $\left\langle\eta(t)\right\rangle=0$, and delta-like correlations $\left\langle \eta(t)\eta(t^{\prime})\right\rangle=2\gamma k_B T \Delta t \delta (t-t^{\prime})$ with Boltzmann's constant $k_B$ and temperature $T$. The Langevin equation can be viewed as resulting from coupling the system to a heat bath with temperature $T$ that causes friction and random forces related by the fluctuation-dissipation theorem. The strength of the coupling to the heat bath is controlled by the friction constant $\gamma$ and for $\gamma=0$ the Langevin equation reduces to Newton's equations of motion.  

The Langevin equation ~(\ref{eq:Langevin}) is solved numerically in small time steps $\Delta t$ using a velocity Verlet-like integration scheme,\cite{Gronbech-Jensen2013} 
%and noise in the form of a Gaussian random number satisfying $\left\langle\eta(t)\right\rangle=0$ and $\left\langle \eta(t)\eta(t^{\prime})\right\rangle=2\gamma k_B T \Delta t \delta (t-t^{\prime})$ with Boltzmann's constant $k_B$, temperature $T$, and time step $\Delta t$.\cite{Gronbech-Jensen2013} The implementation of these dynamics is achieved by modifying the Velocity Verlet algorithm such that the equations of motion correspond to Eq.~(\ref{eq:Langevin}), namely by the integration scheme\cite{Gronbech-Jensen2014}
%
\begin{equation}
    \label{eq:LangevinX}
        \begin{split}
        \mathbf{x}(t+\Delta t)= &\; \mathbf{ x}(t)+b{\mathbf{v}}(t)\Delta t -\frac{b}{2m}\mathbf{\nabla}E(\mathbf{x}(t))\Delta t^2 +\frac{b}{2m}\mathbf{\xi}(t)\Delta t,\\
        \mathbf{v}(t+\Delta t)= &\; a \mathbf{v}(t)-\frac{\Delta t}{2m} \left[ a \mathbf{\nabla} E(\textbf{x}(t))+\mathbf{\nabla} E(\mathbf{x}(t+\Delta t))\right] + \frac{b}{m}\mathbf{\xi}(t+\Delta t),
        \end{split}
\end{equation}
where $b=\left( 1+\frac{\gamma \Delta t}{2m} \right)^{-1}$, and $a=\left( 1-\frac{\gamma \Delta t}{2m} \right) \left( 1+\frac{\gamma \Delta t}{2m} \right)^{-1}$. In the above equations, $\xi(t)$ and $\xi(t+\Delta t)$ denote independent Gaussian random numbers with zero mean and variance $2\gamma k_B T \Delta t$.

This methodology allows SHARC simulations to follow stochastic dynamics and is, therefore, ideally suited for combination with the FFS algorithm. 

%Combining these two methods yields NAFFS, a conceptually novel powerful tool to investigate rare nonadiabatic reactions covering time scales that far exceed those of simulations currently possible. 
%We emphasize that without loss of generality, our approach works independently of the method used to obtain the interatomic interactions that are described by the PESs yielding the potential energy $E$ of an atomic configuration, \textit{i.e.}, it is universally applicable to different levels of theory. In this work, we use analytically constructed models for methodological verification, however, for arbitrary systems, any conceivable method can be chosen.

\section{Implementation}
\label{sec:implementation}

%The combination of FFS to capture rare events and TSH to allow nonadiabatic excited state MD simulations results in NAFFS ---a conceptually novel powerful tool to investigate rare nonadiabatic reactions covering time scales that far exceed those of TSH simulations currently possible. 
%
%In this work, we bring together the FFS algorithm for investigating rare events and a TSH algorithm for carrying out the required excited-state MD simulations in order to allow the investigation of nonadiabatic systems evolving on multiple PESs. 
%In order to implement a methodological realization of this approach, we make use of two established programs which we combine and extend to our intended purpose. 
%In particular, we employ a software capable of performing excited-state MD simulations and another one designed for applying transition path sampling methods.
%
%For the former, we pick SHARC\cite{Mai2019} (Surface Hopping including ARbitrary Couplings), a program package that uses an eponymous approach\cite{Richter2011,Mai2018,Mai2015} for nonadiabatic TSH dynamics simulations. To account for the stochastics of the time propagation as required by the FFS method, we add a Langevin thermostat to the SHARC program, which we implement as described in Sec.~\ref{sec:TSH}.

\begin{figure}
    \centering
    \includegraphics[height=5.0cm]{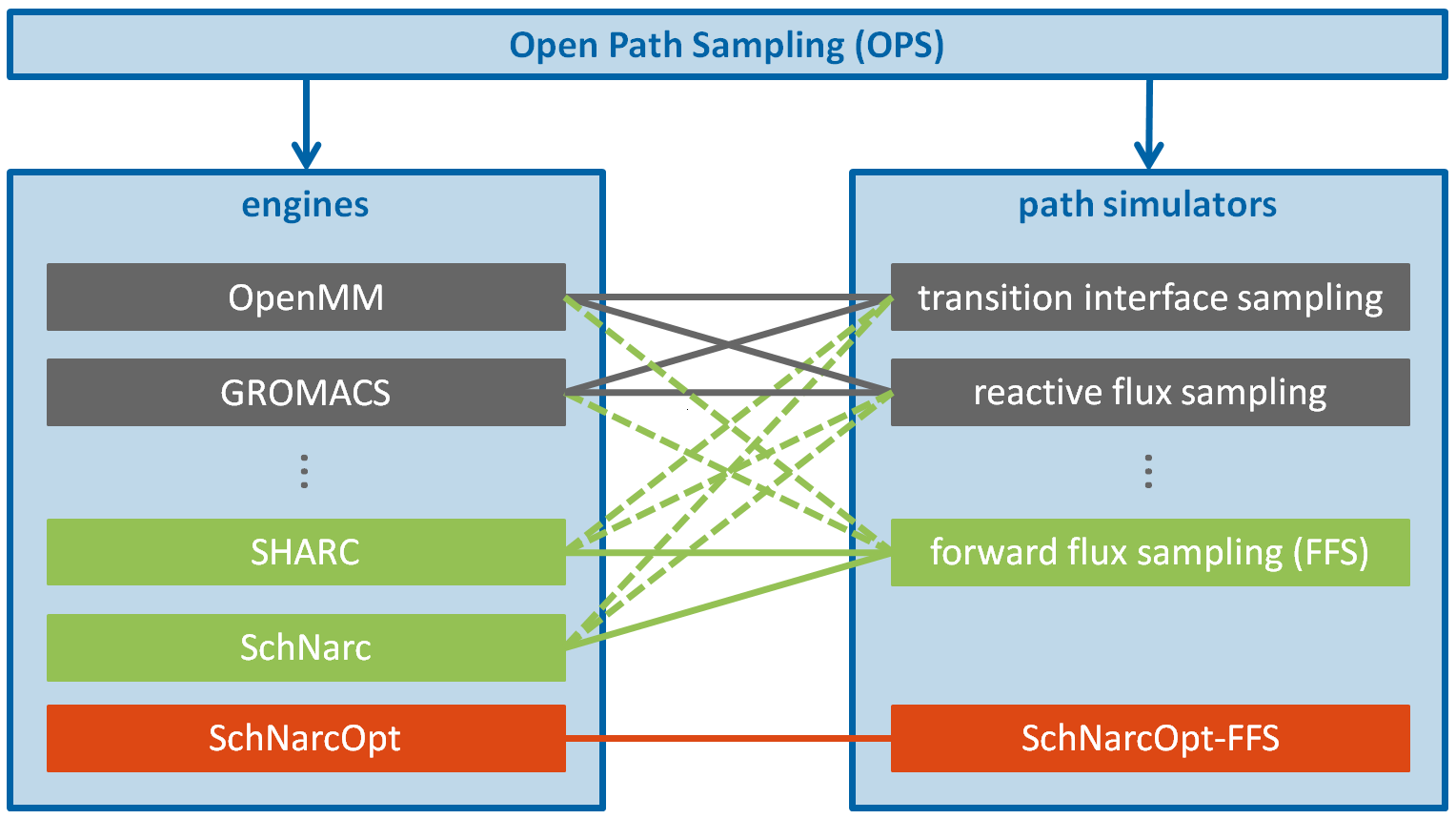}
    \caption{Architecture of the OPS library. Examples of existing engines, simulators, and combinations between them are shown in gray. Green boxes indicate new implementations. Dashed green lines indicate new possible combinations of engines and simulators and solid lines explicit new implementations of NAFFS.
    %(we have not used dashed so far). 
    A dedicated \texttt{SchNarcOpt} engine tailored to a new FFS implementation \texttt{SchNarcOpt-FFS} for computational efficiency is highlighted in red.
    %Architecture of the OPS library with integrated MD engines that can be arbitrarily combined with path simulators. Examples of existing engines and simulators are shown in gray. Green boxes indicate new implementations compatible with any other package in OPS. Green lines indicate possible excited-state combinations (dashed lines have not been used so far). A dedicated \texttt{SchNarcOpt} engine tailored to an FFS implementation \texttt{SchNarcOpt-FFS} for computational efficiency is highlighted in red.
    %Integration of our implementation into the OPS library, which consists of arbitrarily combinable engines (MD programs) and path simulators (TPS methods). Examples of existing ones are shown in gray. Green boxes mark our general implementations in the spirit of OPS, and green lines show possible excited-state combinations. In this work, we apply a dedicated \texttt{SchNarcOpt} FFS implementation tailored to an optimized engine for computational efficiency, shown in red.
    }
    \label{fgr:implementation}
\end{figure}

The practical implementation of NAFFS relies on two program suites. On the TSH side, we extended SHARC\cite{Richter2011,Mai2018,Mai2015} 
with a Langevin thermostat to endow the dynamics with the stochasticity required for FFS  (Sec.~\ref{sec:TSH}).
On the FFS side, we used Open Path Sampling (OPS),\cite{Swenson2019,Swenson2019a} a Python library for path sampling simulations capable to work with various MD codes. 
For example, OPS is interfaced with two popular MD engines: OpenMM\cite{Eastman2010,Eastman2013} and GROMACS\cite{VanDerSpoel2005,Hess2008} (see gray engines in Fig.~\ref{fgr:implementation}).
We have now implemented the SHARC suite and the SchNarc\cite{Westermayr2020} method via the SHARC driver pySHARC\cite{Plasser2019} as new general engines in OPS, see green engines in Fig.~\ref{fgr:implementation}.
Hence, any quantum chemical method compatible with the SHARC program for computing PESs and electronic properties is now also available for NAFFS simulations.
SchNarc was originally developed as an interface between the SHARC program and an extension of the neural network potential SchNet\cite{Schutt2017,Schutt2018} to excited-state properties.\cite{Westermayr2020} 
We do not use neural networks in this work, but because SchNarc is not based on file I/O it allows for computationally much more efficient path sampling simulations using TSH dynamics. 
An additional advantage of the SchNarc engine is that it opens the possibility of using neural network potentials to compute PESs in the future, further decreasing computational costs. 

%In OPS any implemented engine can be combined with any implemented TPS simulator. 
%

To accompany both engines, SHARC and SchNarc, we implemented an OPS tool to capture snapshots, \textit{i.e.}, individual frames of a trajectory.
Such a tool is needed to describe the system at a particular point in time, where specific ranges of PESs are included in addition to the atomic positions and velocities that define the collective variables. 
Both, the ranges of PESs and collective variables, are used for the definition of the stable initial and final regions $A$ and $B$ as well as the definition of the FFS interfaces.
We note that, as usual in OPS, we also include in our engines an additional rejection criterion for discarding trial shots if those  exceed a user-given number of time steps. 
Although this number should be chosen such that (almost) no trajectories are discarded,
such implementation is useful, as it aids, for example, the discovery of new stable minima in the region between $A$ and $B$. 
From a practical point of view, a rejection criterion is also helpful to avoid very long unnecessary calculations in such \textit{a priori} unknown stable minima by terminating the respective trajectory calculation and continuing with the next shot.

In OPS any engine can be combined with any TPS simulator, such as transition interface sampling\cite{VanErp2003} or reactive flux sampling,\cite{VanErp2005} see Fig.~\ref{fgr:implementation}.
As an additional transition path sampling method in the OPS library, here we implemented a general FFS simulator applicable to excited states, see green simulator in Fig.~\ref{fgr:implementation}. 

The advantage of both the general SHARC engines and the general FFS simulator is that they are implemented in the spirit of OPS, \textit{i.e.}, any engine can be combined with any path simulators, so that, in principle, now SHARC (or SchNarc) can be used with any method available in OPS (dashed green lines) and certainly both engines effectively work with FFS (solid green lines).
Despite their flexibility, we have also created a dedicated \texttt{SchNarcOpt} engine that is combined with an optimized \texttt{FFS-SchNarcOpt} path simulator (red in Fig.~\ref{fgr:implementation}), which, although specific, is computationally more efficient since it does not require file I/O. This specific NAFFS implementation has been used in the applications below in order to keep the computational cost as low as possible. 

The vertical excitation and the equilibration process (recall Fig.~\ref{fgr:regions}) pose a necessary preliminary step prior to any NAFFS simulation. In this work, these steps are carried out in a plain SHARC/SchNarc-TSH simulation.
Thus, the workflow of a NAFFS simulation begins with the flux simulation and collection of initial shooting points on the boundary of the initial region $A$ (see Fig.~\ref{fgr:FFS}a).
This process is done via the \texttt{SchNarcOpt} engine through OPS and it is followed by the FFS cycles (see Fig.~\ref{fgr:FFS}b-d), executed via the \texttt{SchNarcOpt-FFS} path simulator in OPS.

In summary, the workflow of a typical NAFFS simulation starts with generating initial conditions (\textit{e.g.}, Wigner sampling and vertical excitation, see Fig.~\ref{fgr:regions}), and a relaxation into the initial region, followed by the NAFFS method that generates transition trajectories between the initial and final regions.

To ensure transparency and reproducibility of the obtained results, the code developed is made freely available (see Sec.~S1~\dag).

%In addition to NAFFS calculations, we also carried out reference simulations based on brute-force TSH trajectories. These are performed via the \texttt{SchNarcOpt} engine through OPS.

\section{Results and Discussion}
\label{sec:results}

As a first application of NAFFS, here we employ two dynamically relevant analytical potential energy landscapes that have been constructed to include rare events in different conditions.
We deliberately choose simple analytical models for testing NAFFS rather than a real molecule because they allow for a systematic investigation of different parameters and demonstrate the broad applicability of NAFFS even in extreme situations, regardless of low or high temperature, strong or weak nonadiabaticity. 
Further, our models can be tuned so that NAFFS results can be compared with reference data obtained with plain brute-force TSH simulations under different conditions. 
These TSH calculations have been also performed via the \texttt{SchNarcOpt} engine through OPS.
The future simulation of real molecules is straightforward, \textit{i.e.}, does not require any further implementation, only investing in the calculation of on-the-fly multidimensional PESs at the desired quantum chemical level of theory.

The first system features an avoided crossing between two states, see Sec.~\ref{sec:1D}.
Using this model, we demonstrate the essential functionality of NAFFS by calculating the temperature dependence of the transition rate constant with the energy barrier of the avoided crossing. 
Further, for this model we investigate the influence of nonadiabatic effects on the transition rate constant by varying the gap size between the PESs and thus the rareness of the event. 

%The second system is a model considered two-dimensional in terms of collective variables, that features a conical intersection between two states in the adiabatic representation, showing richer nonadiabatic dynamics (see Sec.~\ref{sec:2D}). 
The second system includes a conical intersection between two states, see Sec.~\ref{sec:2D}.
This model shows richer nonadiabatic dynamics than the one-dimensional avoided crossing and allows us to focus on the dependence of the reaction rate constant 
%and the rate constants with regard to the number of occurred surface hops 
on temperature and to study the contributions of trajectories with different numbers of hops. 

Both models demonstrate that NAFFS yields correct results compared to reference plain brute-force TSH simulations in a fraction of the computational time and thus is ideally suited to study general rare nonadiabatic reactions.

\subsection{Rare event dynamics through an avoided crossing}
%Potential energy landscape featuring an avoided crossing}
%Analytical model \textbf{1}: a one-dimensional collective variable}
\label{sec:1D}
%We demonstrate our implementation by applying it to a first model system representing a rare barrier crossing event of a single hydrogen-like atom in an analytical potential energy landscape, including each a ground and an excited state potential. Using this model system, we perform FFS simulations using our nonadiabatic FFS version presented in Sec.~\ref{sec:implementation}. 
%We compare their quantitative and qualitative results to the ones obtained by direct brute-force MD simulations. 
%Furthermore, this approach not only shows that our realization of the presented methodology yields the correct reaction rate constants, but also allows to estimate how much the computation time needed is reduced with FFS compared to straightforward MD to arrive at comparably accurate results. 
%Varying the temperature and the diabatic coupling gives insight into how these parameters fundamentally condition the rate.

We define a three-dimensional potential energy landscape from two diabatic harmonic potentials of the form
\begin{equation}
%    V_{\pm}(x,y,z)=\frac{\epsilon}{x_0^2}\left((x\pm 1.0)^2+20\cdot y^2+20\cdot z^2\right),
    V_{\pm}(x,y,z)=\frac{\epsilon}{x_0^2}\left((x\pm x_0)^2+20 y^2+20 z^2\right),
    \label{eq:1D}
\end{equation}
defined in the Cartesian coordinates $x$, $y$, and $z$. In $x$-direction the potential is bi-stable with minima at $\pm x_0$ separated by a barrier of height $\epsilon$.
The tight harmonic potentials acting on the $y$ and $z$ coordinates 
%in Eq.~(\ref{eq:1D})  
make the system effectively one-dimensional 
%allow to consider the system as containing a one-dimensional collective variable 
along the $x$ coordinate.
In particular, the additive terms in $y$ and $z$ are invariant under a diagonalization of the diabatic Hamiltonian.
However, the dynamics is coupled in all directions due to the velocity rescaling (see Eq.~(\ref{eq:velrescaling})) and decoherence scheme (see Eq.~(\ref{eq:decoherence})).
We use system-specific self-consistent units, \textit{i.e.}, we measure energies in units of $\epsilon$, lengths in units of $x_0$, and masses in units of
$m$, where the mass of our system is $1$ (see Sec.~S2.1~\dag\ and Table~S1~\dag) for the remainder of this Section. Time is measured in units of $\sqrt{m x_0^2/\epsilon}$.
%everything is given in these reduced units and we write down only the numerical values. 
%

\begin{figure*}[h]
\centering
    \includegraphics[height=7cm]{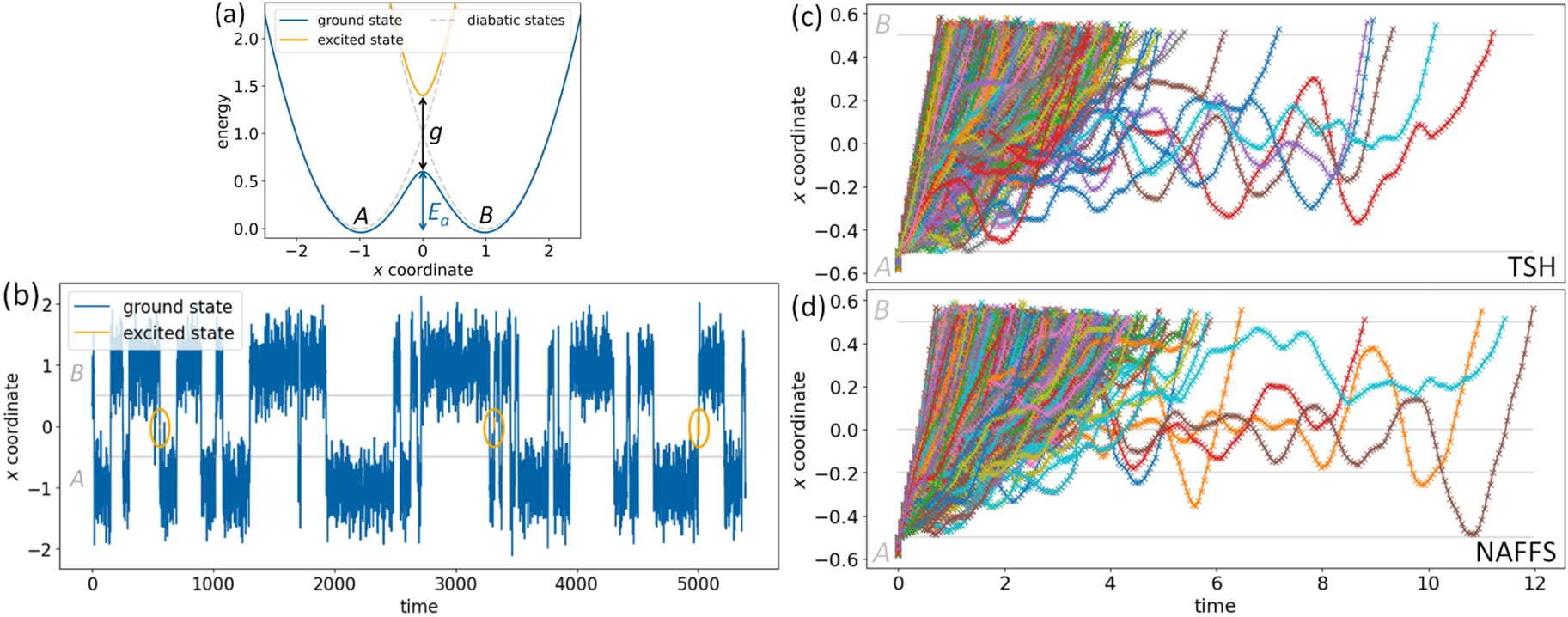}
	\caption{(a) Analytical model system consisting of two diabatic PESs (dashed lines), coupled by a constant coupling of $V_c=0.4$. The corresponding adiabatic (ground and excited) states, shown as solid lines, avoid each other by a gap size $g$. The ground state barrier height is denoted as $E_a$. (b) Representative cutout of a TSH trajectory propagated in this system as a function of time, where the first snapshot of the cutout is shifted to time zero for convenience. Trajectory parts that are spent in the excited state are shown in orange (and additionally marked by orange ovals) \textit{vs.} blue in the ground state. The boundaries of the initial and final regions $A$ and $B$ are depicted in gray. (c-d) Superimposed transition paths between the boundaries of the initial and final regions (gray lines) from TSH (c) and NAFFS (d) simulations within this system with $V_c=0.4$. Different colors allow to discriminate individual transition paths;  symbols indicate calculated time steps. The time of the first snapshot of each transition path is set to zero for convenience.
	The number of plotted transition paths is (c) $1069$ (TSH) and (d) $1803$ (NAFFS). In the latter case, interfaces are also plotted as gray lines. Note that for a better visualization, we only plot transition paths that are shorter than $12$ time units ($>99.5\%$); this leaves out four events in the case of TSH and two in the case of NAFFS (compare with Table~\ref{tab:1D-rates}).}
    \label{fig:1D}
\end{figure*}

The diabatic potentials are coupled to each other by a constant coupling of $V_c$, giving the diabatic Hamiltonian, 
\begin{equation}
    H(x,y,z)=\begin{pmatrix} V_{+}(x,y,z) & V_c \\ V_c & V_{-}(x,y,z) \end{pmatrix},
    \label{eq:1D-HamiltonMatrix}
\end{equation}
which upon diagonalization provides the gap $g$ between the corresponding avoiding adiabatic potentials (here labeled as ground and excited states, see Fig.~\ref{fig:1D}a). 
This model of an avoided crossing is used to compare quantitative and qualitative results against plain brute-force TSH simulations. 
In this way, we demonstrate that NAFFS yields the correct reaction rate constants and evaluate its computational efficiency.  
Varying the temperature and the strength of the diabatic coupling gives insight into how these parameters fundamentally condition the transition rate constants.

For the first simulation we use $V_c=0.4$.  
This is a rather strong diabatic coupling and leads to a very strongly avoided crossing, \textit{i.e.} the energy gap is $g=0.8$ (see Fig.~\ref{fig:1D}a), which is large compared to other choices of the coupling constant considered later. 
Accordingly, one expects very few nonadiabatic surface hops in the generated trajectories and a dynamics that is predominantly adiabatic in the ground state potential with occasional hops to the excited state. 
The ground state resembles a one-dimensional double-well potential, where we define the ground state minimum around $x=-1$ as initial region $A$ and the ground state minimum around $x=1$ as final region $B$.
The two stable regions are separated by a ground state energy barrier of $E_a=0.64$, see Fig.~\ref{fig:1D}a.

All simulations presented in this Section employ the default TSH decoherence parameter\cite{Granucci2010} (see Sec.~S2.2~\dag) and $s=25$ substeps (see Eq.~(\ref{eq:propagator})). 
The remaining adjustable parameters are the temperature, the time step, and the friction constant. 
The time step is set to approximately $\Delta t = 0.0539$, such that one oscillation in a ground state minimum consists of at least $20$ time steps, and, hence, the dynamics of the system can be adequately captured. 
The temperature is set to $T\approx 0.2133$, such that the ground state barrier height is $3~k_BT$. This barrier height is sufficiently large to make the barrier crossing a rare event possible and, at the same time, sufficiently low to enable the observation of transition events in brute-force TSH simulations. 

The rate constant of a transitions from $A$ to $B$ depends on the friction constant $\gamma$ that appears in the Langevin equation (see Eq.~(\ref{eq:Langevin})).\cite{Hanggi1990} 
In particular, the rate constant shows a maximum as a function of the friction constant, known as the Kramers turnover.\cite{Hanggi1990,NatNano2017}
For smaller and larger values of $\gamma$ the rate constant decreases: at low friction because of slow energy exchange of the system with the environment and at high friction due to slow diffusion at the barrier top. The maximum transition rate constant is expected to occur for a friction constant at which the energy dissipated as the system crosses the barrier is about $1~k_B T$. For a barrier shaped like an inverted parabola, this conditions implies $\gamma_{\rm max}/m = \omega k_B T / E_a$, where $\gamma_{\rm max}$ is the friction constant at which the turnover occurs and $\omega$ is the frequency of the unstable mode at the barrier top. For the parameters selected here, the turnover friction is $\gamma_{\rm max} \approx 0.94$. Hence, the friction coefficient selected for our simulations, $\gamma = 1.4133$, is slightly higher than the turnover friction.

For the brute-force TSH trajectory, we run $5$ million time steps, and for the analysis, we define the ground state region $x<-0.5$ as the initial region $A$, and the ground state region $x>0.5$ as the final region $B$.
A representative cutout of the TSH trajectory can be seen in Fig.~\ref{fig:1D}b. 
It shows the typical behavior of a rare event, \textit{i.e.}, it oscillates for a long time in one of the two regions $A$ or $B$ before it undergoes a fast transition to the other region, where it oscillates again. As the diabatic coupling is very large, the crossing is strongly avoided and we expect small nonadiabatic effects. 
Indeed, the system spends only very short periods of time in the excited state, see orange circles in Fig.~\ref{fig:1D}b.
The corresponding reaction rate constant for the transition from the initial region $A$ to the final region $B$ obtained with the brute-force TSH is $(8.25\pm0.28)\cdot 10^{-3}$, see  Table~\ref{tab:1D-rates}.
This result nicely agrees with that obtained by the NAFFS simulation $(8.72\pm0.47)\cdot 10^{-3}$, which aside marginal statistical deviations, demonstrates the quantitative accuracy of our implementation.
The computational details for the NAFFS simulation are given in Table~S2~\dag.
\begin{table}[h]
\small
  \caption{\ Rate constants $k_{AB}$, number of sampled transition paths, mean  transition times, average numbers of hops in adiabatic representation, and the respective standard deviations (std) obtained by brute-force TSH \textit{vs.} NAFFS simulations on the model featuring an avoided crossing, 
  %PES landscape considered in a one-dimensional collective variable 
  with a constant coupling of $V_c=0.4$.}
  \begin{tabular*}{0.48\textwidth}{@{\extracolsep{\fill}}lll}
    \hline
    & TSH & NAFFS \\
    \hline
    %rate constant (ps$^{-1}) & $(1.78\pm 0.06)$ & $(1.88\pm 0.10)$ \\
    rate constant ($10^{-3}$) & $8.25\pm 0.28$ & $8.72\pm 0.47$ \\
    number of transition paths & $1073$ & $1805$ \\
    mean transition time & $2.2$ & $2.1$\\
    std transition time & $2.0$ & $1.0$\\
    mean number of hops & $0.0298$ & $0.0321$ \\
    std number of hops & $0.2424$ & $0.2515$ \\
    \hline
  \end{tabular*}
  \label{tab:1D-rates}
\end{table}
%
% \begin{table}[h]
% \small
%   \caption{\ Computational details for the NAFFS simulation presented in Table~\ref{tab:1D-rates}. %The flux and crossing probabilities are calculated as described in Sec.~\ref{sec:FFS}
%   }
%   \label{tbl:1D-compdetails}
%   \begin{tabular*}{0.48\textwidth}{@{\extracolsep{\fill}}llll}
%     \hline
%     flux sim. & time steps & crossing events & flux ($10^{-3}$) \\
%     \hline
%     & $1$~million & $2137$ & $(77.1\pm 1.7)$ \\
%     \hline
%     NAFFS sim. & interfaces in $x$ & shots per interface & crossing prob. \\
%     \hline
%     & $-0.5\rightarrow -0.2$ & $2000$ & $(24.6\pm1.0)\%$ \\
%     & $-0.2\rightarrow 0.0$ & $2000$ & $(50.9\pm1.2)\%$ \\
%     & $0.0\rightarrow 0.5$ & $2000$ & $(90.25\pm0.67)\%$ \\
%     \hline
%   \end{tabular*}
% \end{table}

Figures~\ref{fig:1D}c and \ref{fig:1D}d show the superimposed transition paths connecting the regions $A$ and $B$, as obtained from TSH and NAFFS  simulations.
The mean time that a transition path takes to go from $A$ to $B$ and the average number of hops occurring during the transition paths are also collected in Table~\ref{tab:1D-rates}, together with their corresponding standard deviations. 
The larger standard deviation for the transition time in TSH compared to NAFFS stems from the overall lower number of transition paths obtained with  TSH and some outlying long transition paths among them.
The good qualitative and quantitative agreement between NAFFS and TSH values confirms that the NAFFS simulation correctly samples transition paths.

% \begin{table}[h]
% \small
%   \caption{\ Mean lengths, average number of hops in adiabatic representation, and the respective standard deviations (std) of transition paths obtained with TSH and NAFFS simulations for the one-dimensional model system with a constant coupling of $V_c=0.4$}
%   \label{tab:1D-lengthshops}
%   \begin{tabular*}{0.48\textwidth}{@{\extracolsep{\fill}}lllll}
%     \hline
%      & mean length & std length & mean hops & std hops \\
%     \hline
%     TSH & $2.2$ & $2.0$ & $0.0298$ & $0.2424$ \\
%     NAFFS & $2.07$ & $0.92$ & $0.0321$ & $0.2515$ \\
%     %TSH & $40.63$ & $35.62$ & $0.0298$ & $0.2424$ \\ in time steps
%     %NAFFS & $38.40$ & $16.90$ & $0.0321$ & $0.2515$ \\ in time steps
%     \hline
%   \end{tabular*}
%   \end{table}

\begin{figure}[h]
\centering
    \includegraphics[height=7.3cm]{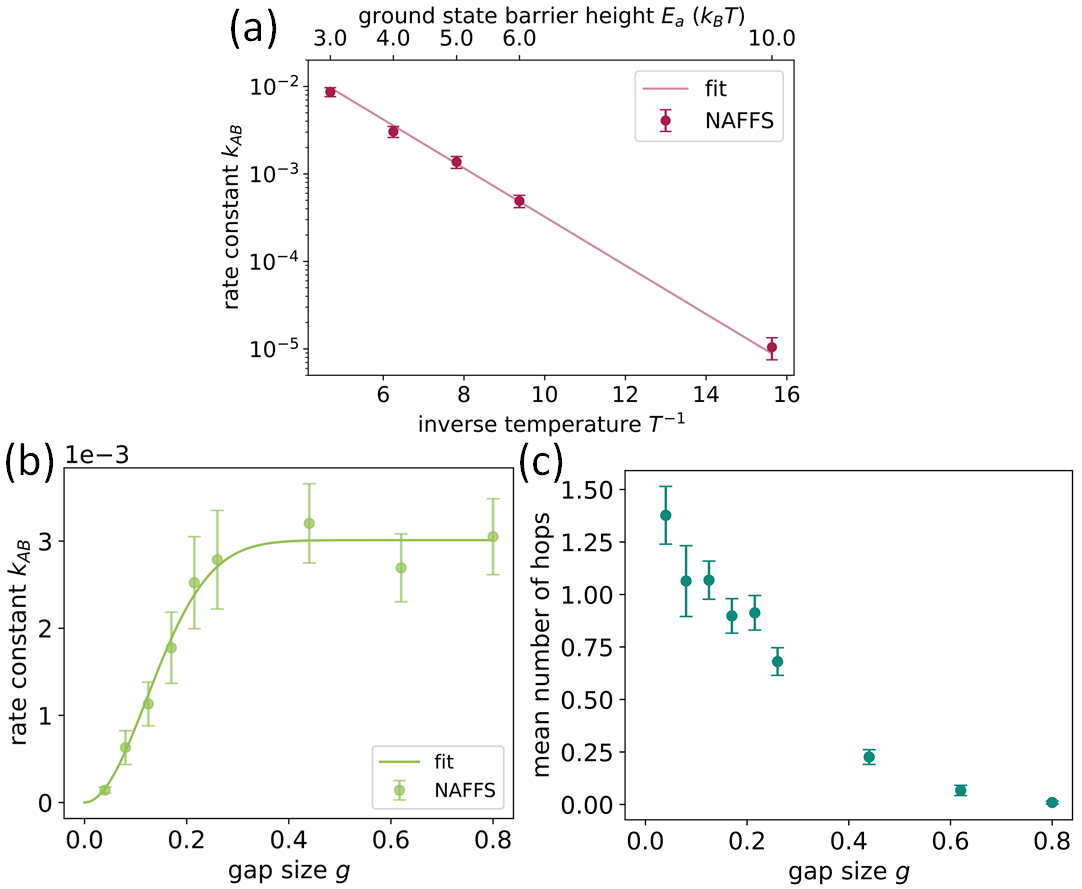}
	\caption{(a) Rate constant $k_{AB}$ of the model system shown in Fig.~\ref{fig:1D}a with $g=0.8$ as a function of the inverse temperature. The scale on the top shows the corresponding barrier height in units of $k_BT$. The linear fit takes only the intercept as a fitting parameter, yielding $\nu=0.19$ (see Eq.~(\ref{eq:Arrhenius})). Error bars are plotted as $2\sigma$ confidence intervals. (b) Rate constant $k_{AB}$ in the system shown in Fig.~\ref{fig:1D}a \textit{vs.} the adiabatic gap size $g$ obtained from NAFFS calculations. Error bars are plotted as $2 \sigma$ confidence intervals. The data are fitted according to Eq.~(\ref{eq:gapsize}), yielding $k_{AB}^0=3.01\cdot 10^{-3}$ and $z=135.72$. (c) Mean number of hops in the adiabatic representation \textit{vs.} adiabatic gap size as obtained from NAFFS transition paths. The obtained standard deviations of the mean are plotted as error bars.}
    \label{fgr:1D-ArrheniusGapsize}
\end{figure}

In order to demonstrate the applicability of NAFFS in harsher conditions, we now decrease the temperature, which effectively increases the barrier and thus the rareness of the event.
To avoid variations of the rate constant at different temperatures due to the definition of the boundaries of the stable regions $A$ and $B$, we fix $A$ as $x<-1$ 
and $B$ as $x>1$ on the ground state for the remaining simulations on this model.  
We expect the temperature dependence of calculated rate constants to follow Arrhenius' law,
\begin{equation}
    k_{AB}=\nu \exp{\left( -\frac{E_a}{k_B T}\right) },
    \label{eq:Arrhenius}
\end{equation}
with prefactor $\nu $. 
%Boltzmann's constant $k_B$, temperature $T$, and activation energy $E_a$, where the latter is given by the barrier height of the system (recall Fig.~\ref{fig:1D}a). 
The reaction rate constants obtained with NAFFS, shown in Figure~\ref{fgr:1D-ArrheniusGapsize}a as a function of the inverse temperature, fit Arrhenius' law remarkably well. 

Table~\ref{tab:compsavings} illustrates the computational efficiency of NAFFS against TSH, showing that the speedup of NAFFS versus TSH increases with the rareness of the transition.

Note that the average number of time steps needed to obtain one NAFFS trajectory depends on the acceptance probabilities for the various interfaces (especially of the last shooting interface where the total number of final transition paths is determined), and, hence, on the choice of the interface placements. 
This explains that the average number of time steps required to obtain one reactive NAFFS path is always of the same order of magnitude, in stark contrast to the TSH simulations, which require an increasingly larger number of steps with decreasing temperature. 
%However, in contrast to brute-force TSH calculations, its order of magnitude does not change with an increasing barrier height. %
Accordingly, for the lowest temperature (largest barrier height $10~ k_B T$), NAFFS sampled $876$ transition paths in about one million time steps (flux calculation included) whereas TSH sampled only $4$ in the same amount of time steps.
This implies a speed up of almost $200$ in favor of NAFFS, \textit{i.e.}, a sampling acceleration of two orders of magnitude. 
We note that at this low temperature, an accurate rate constant calculation is no longer possible within reason with brute-force TSH, but well feasible with NAFFS.
\begin{table}[h]
\small
  \caption{\ Average number of time steps needed to obtain one transition path in TSH and in NAFFS  (flux simulation included) calculations, for different ground state barrier heights $E_a$ (see Fig.~\ref{fgr:1D-ArrheniusGapsize}a). The speedup factor in favour of NAFFS over TSH is also shown.}
  \label{tab:compsavings}
  \begin{tabular*}{0.48\textwidth}{@{\extracolsep{\fill}}llll}
    \hline
    $E_a~(k_BT)$ & TSH (time steps) & NAFFS (time steps) & speedup factor \\
    \hline
    $ 3.0 $ & $ 2464 $ & $ 1083 $ & $ 2.3 $ \\
    $ 4.0 $ & $ 7219 $ & $ 2459 $ & $ 2.9 $ \\
    $ 5.0 $ & $ 16 677 $ & $ 1750 $ & $ 9.5 $ \\
    $ 6.0 $ & $ 38 889 $ & $ 1120 $ & $ 34.7 $ \\
    $ 10.0 $ & $ 226 465 $ & $ 1228 $ & $ 184.4 $ \\
    \hline
  \end{tabular*}
\end{table}

Finally, since the dynamics with $V_c=0.4$ is predominantly adiabatic in the ground state potential, we investigate the effect of varying the diabatic coupling from $0.4$ to $0.02$, thus decreasing the diabatic gap from $0.8$ to $0.04$, and thus increasing the nonadiabaticity of the avoided crossing. 
The barrier height in terms of $k_B T$ is kept constant and equal to $4~k_B T$ for all simulations. 
The rate constant is expected to decrease with smaller gaps, as the nonadiabatic effects increase, \textit{i.e.}, the number of hops in the adiabatic representation increases, and, thus, the probability of a transition for a trajectory that reaches the energy barrier is lower than for systems with higher gap sizes. 
In other words, with smaller gaps, the nonadiabatic effects become more an obstacle that the system must overcome to complete a transition in addition to the potential energy barrier. In the limit of no diabatic coupling ($V_c=0$ and, hence, $g=0$), the system is diabatically trapped, \textit{i.e.}, the dynamics purely evolves on one diabatic state, and the transition rate constant is zero.

As expected, the calculated reaction rate constants are lower for smaller gap sizes, see  Figure~\ref{fgr:1D-ArrheniusGapsize}b.
%Figure~\ref{fgr:1D-ArrheniusGapsize}b shows the dependence of the reaction rate constant on the adiabatic gap size, which is lower for smaller gap sizes due to nonadiabatic effects.
The probability that a trajectory coming from the initial region $A$ hops to the excited state in the vicinity of the barrier, oscillates there for one period and then falls back to the ground state in direction of $A$ due to the inertia, increases with decreasing gap size, \textit{i.e.}, the closer the adiabatic states come to each other. 
Accordingly, the mean number of surface hops in transition paths also increases with decreasing gap size, see Fig.~\ref{fgr:1D-ArrheniusGapsize}c. 
%Since this nonadiabatic effects reduce the probability of the system of transitioning to the final region $B$, the reaction rate constant is lower for lower gap sizes (see Fig.~\ref{fgr:1D-ArrheniusGapsize}b). 
%
We fitted the obtained rate constants (see Fig.~\ref{fgr:1D-ArrheniusGapsize}b) according to the Landau-Zener-type formula\cite{Novaro2012,Zener1932,Landau1932}
\begin{equation}
    k_{AB}(g)=k_{AB}^0 \cdot \left( 1-e^{ -z \cdot g^2} \right)
    \label{eq:gapsize}
\end{equation}
where  $k_{AB}^0$ is the ground state transition rate constant, and $z$ is a fitting parameter $z$. This expression approximately describes the dependence of the reaction rate constant on the gap size.\cite{Hanggi1990} 
As shown in Fig.~\ref{fgr:1D-ArrheniusGapsize}b, the fit nicely reproduces our data.
Even for small energy gap sizes, and, thus, highly nonadiabatic situations, the NAFFS and TSH simulation rate constants agree (see Fig.~S1~\dag), further validating our method.

%The presented results sufficiently prove that our method provides quantitatively and qualitatively correct results for ground state as well as excited-state systems and is hence generally applicable. 

%In the following Sec.~\ref{sec:2D} we present a second analytical model system defined as a two-dimensional PES landscape in terms of collective variables that follows inherently nonadiabatic dynamics. Using two dimensions enriches the dynamics by several aspects that one-dimensional models cannot provide, such as describing conical intersections.

\subsection{Rare event dynamics through a conical intersection}
%Potential energy landscape featuring a CI
%\subsection{Analytical model \textbf{2}: two-dimensional collective variables}
\label{sec:2D}
%We present our method applied to a second model system in an analytical potential energy landscape whose dynamics are of strong nonadiabatic nature. With this system, we carry out and compare brute-force TSH and NAFFS simulations using our nonadiabatic FFS method introduced in Sec.~\ref{sec:implementation}. This allows to illustrate the capability of our NAFFS method of sampling rare transition paths in excited-state systems that involve multiple PESs featuring conical intersections.

\begin{figure*}[h]
\centering
    \includegraphics[height=8cm]{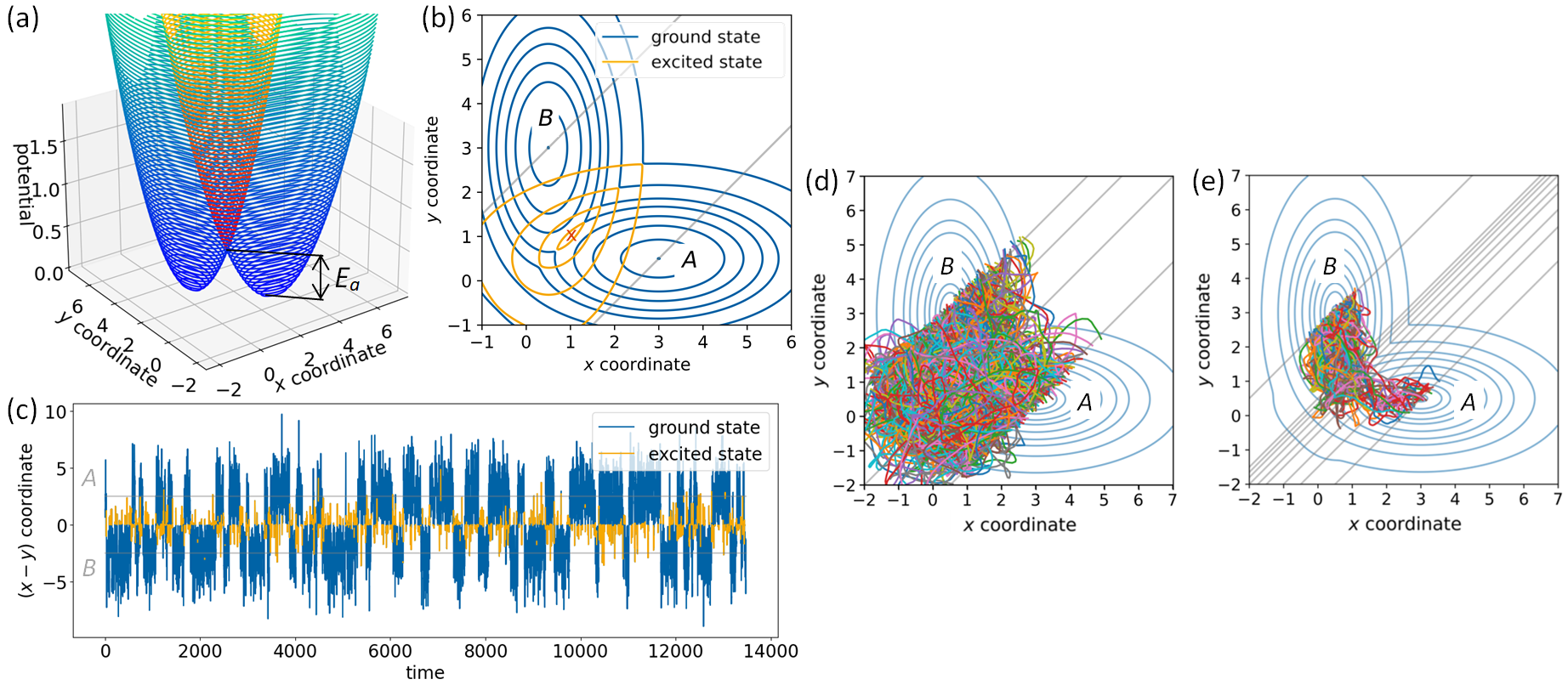}
    \caption{(a-b) Adiabatic representation of the potential energy landscape of Eq.~(\ref{eq:2D_11})-(\ref{eq:2Dcoupling}), shown as a function of $x$ and $y$ as a surface in three dimensions (a) and as a contour plot, where the boundaries of the initial and final regions $A$ and $B$ are shown in gray (b). The ground state is plotted in blue and the excited state in orange. In (b), the location of the conical intersection is marked in red. (c) Representative cutout of a TSH trajectory projected on $(x-y)$ obtained for the model system as a function of time, where the first snapshot of the cutout is shifted to time zero for convenience. For better overview, stable state boundaries are shown in gray. Sojourns in the ground state are plotted in blue, stays in the excited adiabatic state are plotted in orange. (d-e) Transition paths obtained in the NAFFS simulation for temperatures chosen such that the ground state barrier height equals $1~k_B T$ (d) and $6~k_B T$ (e). For the latter the distribution is narrower than for the former, where higher energy regions of the PESs can be visited. Adiabatic ground state contour lines are shown in blue, and interfaces used in the NAFFS simulations in gray.}
    \label{fig:2D}
\end{figure*}

To examine the application of NAFFS to rare event dynamics in the vicinity of an explicit conical intersection, we consider a model with two coupled diabatic potential energy surfaces,
\begin{equation}
    V_{11}(x,y)=a (x-c)^2 +b (y-d)^2+e z^2
    \label{eq:2D_11}
\end{equation}
and
\begin{equation}
    V_{22}(x,y)=b (x-d)^2 +a (y-c)^2+e z^2
    \label{eq:2D_22}
\end{equation}
with Cartesian coordinates $x$, $y$, and $z$ and parameters $a=0.512$, $b=0.128$, $c=0.5$, $d=3.0$, $e=12.8$.
The narrow harmonic potential around the $z$ coordinate allows us to consider the potential energy landscape as a function of two variables, $x$ and $y$.
Again, all values are given in system-specific self-consistent units (see Sec.~S2.1~\dag~and Table~S1~\dag) and the mass is $m=1$. 
The coupling between the two diabatic PESs is given by
\begin{equation}
    V_{12}(x,y)=V_{21}(x,y)=k(x+y-f)
    \label{eq:2Dcoupling}
\end{equation}
with the prefactor $k=0.0128$, and $f=2.3$, resulting in the adiabatic PESs shown in Fig.~\ref{fig:2D}a-b. 
%By definition, the energy landscape contains a true conical intersection, which additionally entails richer dynamics compared to the system presented in Sec.~\ref{sec:1D}.

We choose the default decoherence parameter \cite{Granucci2010} (see Sec.~S2.3~\dag) and $s=25$ substeps (see Eq.~(\ref{eq:propagator})).
As meaningful parameter values for the propagation using the Langevin thermostat, we choose a time step of $0.1348$, a temperature $T\approx 0.6370$, and a friction coefficient $\gamma\approx 0.7995$.
In contrast to the model discussed in Sec.~\ref{sec:1D}, here the rare event is not due to a high barrier, as this is only $1~k_BT$, but due to a small diabatic coupling 
%(see Eq.~(\ref{eq:2Dcoupling})) 
that makes the system stay preferably on one \textit{diabatic} surface (\textit{i.e.}, diabatic trapping).
Accordingly, the dynamics shows frequent hops between the ground and excited adiabatic PESs in regions close to the conical intersection (see Fig.~\ref{fig:2D}c), \textit{i.e.}, in general twice during a typical oscillation period in one of the minima.
Hence, in this case the rare event is a rare hop in the diabatic representation, \textit{i.e.}, a transition between the two diabatic PESs $V_{11}$ and $V_{22}$ (Eq.~(\ref{eq:2D_11})-(\ref{eq:2D_22})).
The initial region $A$ and final region $B$ are defined by $(x-y)\geq 2.5$ and $(x-y)\leq -2.5$, respectively, plus the additional condition that the system needs to be located in the ground state.
The stable region's boundaries also correspond to the first and last interface in the NAFFS simulation.

For this model, we performed a plain brute-force TSH simulation of $5$ million time steps. The resulting rate constant, $(5.58 \pm 0.13)\cdot 10^{-3}$, for the transition from $A$ to $B$ agrees very well with the rate constant obtained using a NAFFS calculation, ($5.80 \pm  0.30)\cdot 10^{-3}$, performed with $1$ million time steps in the flux simulation followed by $2000$ shots per interface  (see Table~\ref{tbl:2D-rates}). Computational details for the NAFFS simulations are given in Table~S3 of Sec.~S2.2~\dag.
A second NAFFS simulation of half the size of the previous one ($0.5$ million time steps and $1000$ shots per interface, see Sec.~S2.3~\dag) still yields the correct result, namely $k_{AB}=(5.56 \pm 0.41)\cdot 10^{-3}$, highlighting the efficiency of NAFFS. 
\begin{table}[h]
\small
  \caption{\ Rate constant $k_{AB}$, number of sampled reactive paths, average transition times, average number of hops in adiabatic representation, and the respective standard deviations (std) obtained by brute-force TSH \textit{vs.} NAFFS simulations, for the model with a conical intersection}
  \label{tbl:2D-rates}
  \begin{tabular*}{0.48\textwidth}{@{\extracolsep{\fill}}lll}
    \hline
    & TSH & NAFFS \\
    \hline
    rate constant ($10^{-3}$) & $5.58 \pm 0.13$ & $5.80 \pm  0.30$ \\
    number of transition paths & $1857$ & $1025$ \\
    mean transition time & $87.67$ & $87.98$ \\
    std transition time & $47.30$ & $45.46$ \\
    mean number of hops & $2.89$ & $2.73$ \\
    std number of hops & $2.10$ & $2.13$ \\
    \hline
  \end{tabular*}
\end{table}
%

\begin{figure}[h]
\centering
    \includegraphics[height=8.3cm]{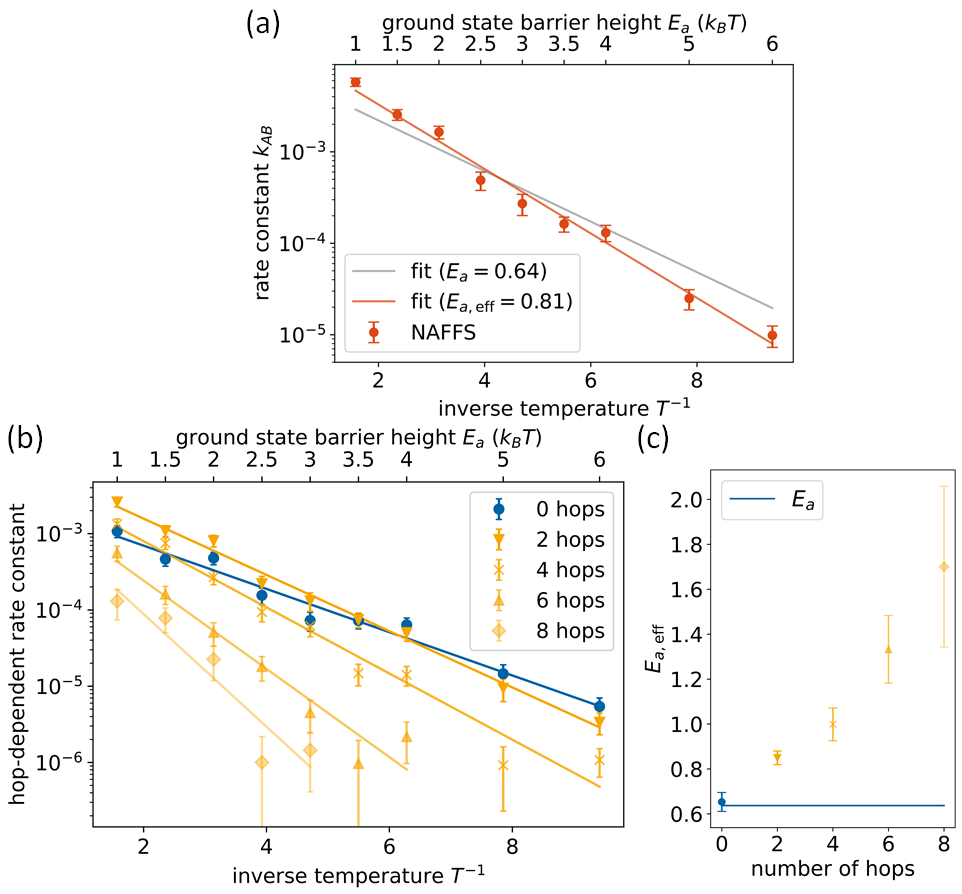}
	\caption{(a) Rate constant $k_{AB}$ as a function of the inverse temperature (the corresponding ground state barrier in units of $k_{\rm B}T$ in indicated on the upper axis). Error bars are plotted as $2\sigma$ confidence intervals. The data are fitted according to Eq.~(\ref{eq:Arrhenius-eff}), yielding $E_{a,\mathrm{eff}}=0.812$ and $\nu=0.017$, and to Eq.~(\ref{eq:Arrhenius}), yielding $\nu=0.0079$. (b) Reaction rate constants due to transition paths from $A$ to $B$ that have a certain number of hops, as a function of the temperature. Blue indicates a transition path purely in the adiabatic ground state (zero hops) and orange implies transition paths involving the excited state. The data are fitted according to Eq.~(\ref{eq:Arrhenius-eff}). Data for more than $8$ hops  (see Fig.~S2c~\dag) are not statistically significant. 
	(c) Effective activation energies as a function of the number of hops. Blue line indicates the ground state energy barrier. Error bars are shown as $2\sigma$ confidence intervals.}
    \label{fgr:2D-ArrheniusHops}
\end{figure}

%Qualitatively, 
The transition paths obtained by the NAFFS simulations (Fig.~\ref{fig:2D}d) are very similar to the ones obtained from brute-force TSH simulations (Fig.~S2a~\dag), demonstrating that NAFFS correctly samples transition paths in strong nonadiabatic regimes. 
Therefore, we next change the parameters of our model system to study it under different conditions.
First, we investigate the dependence of the reaction rate constant on temperature.
As can be seen in Fig.~\ref{fgr:2D-ArrheniusHops}a, due to the stronger nonadiabaticity of the system, the dependence is stronger than in the case of the avoided crossing (recall Fig.~\ref{fgr:1D-ArrheniusGapsize}a), \textit{i.e.}, the slope in a $\log(k_{AB})$ \textit{vs.} $T^{-1}$ plot is steeper than that given by Arrhenius' law (Eq.~(\ref{eq:Arrhenius})) with an activation energy that equals the ground state barrier height (see Fig.~\ref{fgr:2D-ArrheniusHops}a). Fitting the reaction rate constants with the expression %
\begin{equation}
    k_{AB}=\nu \exp{\left( -\frac{E_{a,\text{eff}}}{k_B T}\right) },
    \label{eq:Arrhenius-eff}
\end{equation}
with constant $\nu$ yields an effective activation energy of $E_{a,\text{eff}}=0.812\pm 0.036$, which is significantly higher than the ground state energy barrier of $E_a=0.64$. This means that the nonadiabatic effects lead to an additional barrier that decreases the probability of the system to undergo a transition.

%This allows us to conclude that excited-state rare systems, whose rareness is significantly influenced by nonadiabatic effects, require an effective activation energy in order to undergo a transition. 

%This effective activation energy is higher than the potential energy barrier of the adiabatic ground state, \textit{i.e.}, nonadiabatic effects represent an additional barrier that decreases the probability of the system of undergoing a transition.In our model, the effective activation energy can be obtained fitting our data according to
%
%\begin{equation}
%k_{AB}=\nu \cdot \exp{\left( -\frac{E_{a,\text{eff}}}{k_B
%T}\right) },
%    \label{eq:Arrhenius-eff}
%\end{equation}
%
%with constant $\nu$. This yields an effective activation energy of $E_{a,\text{eff}}=0.812\pm 0.036$, which is significantly higher than the ground state energy barrier of $E_a=0.64$.

The obtained NAFFS transition paths show the expected qualitative behavior for different temperatures: their distribution is broader for higher temperatures as the system has more energy available, and is narrower for low temperatures where the transition paths are located in the region around the conical intersection.
In regions close to the conical intersection, the energy that the system needs to transition to the final region $B$ is lower than in regions far away from the conical intersection, see the narrower distribution of transition paths in Fig.~\ref{fig:2D}e along the $x=y$ diagonal direction compared to Fig.~\ref{fig:2D}d.
At low temperatures (see Fig.~\ref{fig:2D}e), the rare event is mainly determined by the high potential barrier, whereas for high temperatures (see Fig.~\ref{fig:2D}d), the rareness is predominantly caused by the nonadiabatic effects.

Figure~\ref{fgr:2D-ArrheniusHops}b shows that the rate constants of reactive paths that exhibit a given number of hops between the ground and the excited adiabatic PESs 
%scaled by the fraction of paths that showed this number of hops to account for the fact that the distribution of the number of hops is different 
differ from each other. 
In general, the fraction of transition paths that do not undergo hops increases with decreasing temperature, while the fraction of transition paths that feature hops decreases with decreasing temperature (see Fig.~S2b~\dag).
This is because the system needs energy to undergo hops to the excited state, but has less energy available the lower the temperature is (see Fig.~S2d~\dag).
Note that due to using velocity rescaling (see Eq.~(\ref{eq:velrescaling}), the number of hops to the excited state might be slightly overestimated, because the system can use all the kinetic energy along the velocity vector to perform a hop.
Rescaling along the NAC direction---usually regarded as more accurate\cite{Subotnik2016}---would lead to fewer upwards hops, because less kinetic energy is usually available along the NAC direction. For simplicity, here we use velocity rescaling.
%Nonadiabatic couplings are not implemented in the analytical models used, which is why we resort to velocity rescaling.
%Therefore, we do not claim to be able to draw conclusions from these results about how much energy a hop costs the system on average, and Fig.~\ref{} could be slightly different for NAC rescaling.
%Nevertheless, the results obtained are in line with expectations considering the dynamics used.
%We emphasize that, unlike the implementation of the analytical models used, SHARC and SchNarc generally work with NACs, and the fact that we performed velocity rescaling here for the analytical models is an artifact of the implementation of the analytical models, not of the method \textit{per se}, and it does not represent a limitation of the latter.
%

Since we defined the initial and final regions $A$ and $B$ in the adiabatic ground state, a transition path from $A$ to $B$ can only have an even number of hops in the adiabatic representation.
The ``hop-dependent'' rate constants (see Fig.~\ref{fgr:2D-ArrheniusHops}b) are fitted according to Eq.~(\ref{eq:Arrhenius-eff}), yielding effective activation energies $E_{a,\text{eff}}$ for transition paths with a certain number of hops between the ground and excited adiabatic PESs.
$E_{a,\text{eff}}$ increases with increasing number of hops, which again agrees with the necessity to spend energy for hopping events (see Fig.~\ref{fgr:2D-ArrheniusHops}c).
The effective activation energy for transition paths showing zero hops, \textit{i.e.}, reaction paths evolving entirely in the ground state, is $E_{a,\text{eff}}=0.653\pm 0.043$, which within the range of uncertainty aligns accurately with the ground state energy barrier of $E_a=0.64$.
Hence, transition paths evolving only in the ground state follow Arrhenius' law even if this ground state is part of a highly nonadiabatic PES landscape.
Furthermore, nonadiabatic transition paths also follow approximately Arrhenius' law, but with a higher effective activation energy. The latter can be understood when thinking about nonadiabatic effects as prolonging the transition path (\textit{e.g.}, because of turning around and having to come back again), which can be compensated by higher energies and hence the reaction barrier seems effectively higher.

An estimate of the computational savings obtained when using our NAFFS implementation is shown in Table~\ref{tab:compsavings_2D}.
As in the case of the rare event dynamics through an avoided crossing (Table~\ref{tab:compsavings}), here we also achieve a speedup factor of about $200$ for the largest barrier height considered.
%and, hence, bridge time scales of two orders of magnitude.

\begin{table}[h]
\small
  \caption{\ Average number of time steps needed in an NAFFS simulation to obtain one reactive path (flux simulation included) and average number of time steps needed in a brute-force TSH simulation to obtain one reactive path, shown for different ground state barrier heights $E_a$ (see Fig.~\ref{fgr:2D-ArrheniusHops}a). The speedup factor in favour of NAFFS over TSH is also shown}
  \label{tab:compsavings_2D}
  \begin{tabular*}{0.48\textwidth}{@{\extracolsep{\fill}}llll}
    \hline
    $E_a~(k_BT)$ & TSH (time steps) & NAFFS (time steps) & speedup factor \\
    \hline
    $ 1.0 $ & $ 2865 $ & $ 1165 $ & $ 2.5 $ \\
    $ 1.5 $ & $ 5618 $ & $ 1040 $ & $ 5.4 $ \\
    $ 2.0 $ & $ 12195 $ & $ 837 $ & $ 14.6 $ \\
    $ 2.5 $ & $ 26316 $ & $ 850 $ & $ 31.0 $ \\
    $ 3.0 $ & $ 66667 $ & $ 732 $ & $ 91.1 $ \\
    $ 3.5 $ & $ 66667 $ & $ 1987 $ & $ 33.6 $ \\
    $ 4.0 $ & $ 166667 $ & $ 1495 $ & $ 111.5 $ \\
    $ 5.0 $ & $ 1000000 $ & $ 8460 $ & $ 118.2 $ \\
    $ 6.0 $ & $ 1000000 $ & $ 4968 $ & $ 201.3 $ \\
    \hline
  \end{tabular*}
\end{table}

%Conceivably, many additional investigations of the nonadiabatic dynamics of our analytical models can be performed and would certainly provide further instructive insights into the nature of these models. We have provided some of these in Sec.~S2.3 in the ESI~\dag. However, we conclude at this point, because the purpose of demonstrating the usefulness, correctness, accuracy, and potential of the excited-state FFS method we have developed and implemented has been sufficiently shown.

% --- CD ---

\section{Conclusions}
\label{sec:conclusions}

In this work, we have introduced a nonadiabatic forward flux sampling (NAFFS) method that uses the trajectory surface hopping (TSH) algorithm for the underlying dynamics simulation. 
This method extends the previous fields of application of FFS to capture rare events in electronically excited systems, 
%by introducing a nonadiabatic treatment of the time propagation, 
such as those initiated by the absorption of a photon. 
NAFFS is therefore suitable to deal with excited-state processes that occur on very long time scales, which cannot be otherwise accessed with plain brute-force TSH simulations.
%on time scales of nanoseconds and beyond accessible --- far larger than currently possible with brute-force calculations even if machine learning potentials are used as a cheaper alternative to quantum chemical calculations while maintaining \textit{ab initio} accuracy.\cite{Westermayr2019}  
Using two models that exemplify different regimes of rareness and nonadiabaticity, we demonstrate that NAFFS produces quantitatively and qualitatively correct results at a computational cost that is two orders of magnitude lower than that of conventional TSH molecular dynamics simulations.
%depending on the rareness of the event. 
Unlike previous efforts to develop nonadiabatic transition path sampling methods,\cite{Sherman2016} our method does not need to propagate trajectories back in time, 
%as many TPS methods other than FFS would require, 
and, hence, avoids serious problems in long simulations that include several hopping events.\cite{Subotnik2015,Schile2018} 
%Furthermore, NAFFS can be applied to arbitrary chemical systems without adapting our implementation.

The presented approach is particularly promising to investigate photoinduced chemical reactions that are hindered by potential energy barriers or very small nonadiabatic couplings, and thus take a long time to occur. 
Exciting examples include DNA damage and repair processes,\cite{Barbatti2015,Improta2016} enone [2+2] photocycloadditions\cite{Brimioulle2013} and many more.
%some of which are of great scientific importance beyond fundamental research for societal challenges. In medicine, for example, the study of photoreactions of thymines in the DNA could help us better understand the origin of skin cancer.\cite{Barbatti2015,Improta2016,Rauer2016,Rauer2018} In the growing research field of sustainable chemistry, the investigation of light-induced pathways such as of pyridones can advance light-driven photochemical synthesis, which is becoming increasingly important.\cite{Ciamician1912,Chen2019,Brimioulle2013}

%In this work, we have presented a method that shows promise for addressing such problems in the future. Currently, we are working on the application of our method to a chemically relevant rare photoreaction.

\section*{Conflicts of interest}
There are no conflicts to declare.

\section*{Acknowledgements}
The authors thank the University of Vienna for continuous support, in particular the one provided in the framework of the research platform ViRAPID. M.~X.~T., and L.~G. appreciate additional support provided by the Austrian Science Fund, W~1232 (MolTag). The computational results presented have been achieved (in part) using the Vienna Scientific Cluster (VSC). The authors thank the SHARC development team, and the ViRAPID members for insightful discussions. We thank Barbara Wagner for her contributions to some NAFFS calculations.

%%%END OF MAIN TEXT%%%


\balance

\renewcommand\refname{References}

%%%REFERENCES%%%
\bibliography{library} %You need to replace ''rsc'' on this line with the name of your .bib file
\bibliographystyle{rsc} %the RSC's .bst file

\end{document}

\pagestyle{fancy}
\thispagestyle{plain}
\fancypagestyle{plain}{
%%%HEADER%%%
\renewcommand{\headrulewidth}{0pt}
}
%%%END OF HEADER%%%

%%%PAGE SETUP - Please do not change any commands within this section%%%
\makeFNbottom
\makeatletter
\renewcommand\LARGE{\@setfontsize\LARGE{15pt}{17}}
\renewcommand\Large{\@setfontsize\Large{12pt}{14}}
\renewcommand\large{\@setfontsize\large{10pt}{12}}
\renewcommand\footnotesize{\@setfontsize\footnotesize{7pt}{10}}
\makeatother

\renewcommand{\thefootnote}{\fnsymbol{footnote}}
\renewcommand\footnoterule{\vspace*{1pt}% 
\color{cream}\hrule width 3.5in height 0.4pt \color{black}\vspace*{5pt}} 
\setcounter{secnumdepth}{5}

\makeatletter 
\renewcommand\@biblabel[1]{#1}            
\renewcommand\@makefntext[1]% 
{\noindent\makebox[0pt][r]{\@thefnmark\,}#1}
\makeatother 
\renewcommand{\figurename}{\small{Fig.}~}
\sectionfont{\sffamily\Large}
\subsectionfont{\normalsize}
\subsubsectionfont{\bf}
\setstretch{1.125} %In particular, please do not alter this line.
\setlength{\skip\footins}{0.8cm}
\setlength{\footnotesep}{0.25cm}
\setlength{\jot}{10pt}
\titlespacing*{\section}{0pt}{4pt}{4pt}
\titlespacing*{\subsection}{0pt}{15pt}{1pt}
%%%END OF PAGE SETUP%%%

%%%FOOTER%%%
\fancyfoot{}
%\fancyfoot[LO,RE]{\vspace{-7.1pt}\includegraphics[height=9pt]{head_foot/LF}}
%\fancyfoot[CO]{\vspace{-7.1pt}\hspace{13.2cm}\includegraphics{head_foot/RF}}
%\fancyfoot[CE]{\vspace{-7.2pt}\hspace{-14.2cm}\includegraphics{head_foot/RF}}
\fancyfoot[RO,LE]{\footnotesize{\sffamily{1--\pageref{LastPage} ~\textbar  \hspace{2pt}\thepage}}}
%\fancyfoot[LE]{\footnotesize{\sffamily{\thepage~\textbar\hspace{2pt} 1--\pageref{LastPage}}}}
\fancyhead{}
\renewcommand{\headrulewidth}{0pt} 
\renewcommand{\footrulewidth}{0pt}
\setlength{\arrayrulewidth}{1pt}
\setlength{\columnsep}{6.5mm}
\setlength\bibsep{1pt}
%%%END OF FOOTER%%%

%%%FIGURE SETUP - please do not change any commands within this section%%%
\makeatletter 
\newlength{\figrulesep} 
\setlength{\figrulesep}{0.5\textfloatsep} 

\newcommand{\topfigrule}{\vspace*{-1pt}% 
\noindent{\color{cream}\rule[-\figrulesep]{\columnwidth}{1.5pt}} }

\newcommand{\botfigrule}{\vspace*{-2pt}% 
\noindent{\color{cream}\rule[\figrulesep]{\columnwidth}{1.5pt}} }

\newcommand{\dblfigrule}{\vspace*{-1pt}% 
\noindent{\color{cream}\rule[-\figrulesep]{\textwidth}{1.5pt}} }

\makeatother
%%%END OF FIGURE SETUP%%%

%%%FONT SETUP - please do not change any commands within this section
\renewcommand*\rmdefault{bch}\normalfont\upshape
\rmfamily
\section*{}
\vspace{-1cm}

\renewcommand{\thesection}{S\arabic{section}}
\renewcommand{\thetable}{S\arabic{table}}
\renewcommand{\thefigure}{S\arabic{figure}}
\renewcommand{\thesubsection}{S\arabic{section}.\arabic{subsection}}
\renewcommand{\thesubsubsection}{S\arabic{section}.\arabic{subsection}.\arabic{subsubsection}}
%%%FOOTNOTES%%%

\footnotetext{\textit{$^{a}$~Research Platform on Accelerating Photoreaction Discovery (ViRAPID), University of Vienna, Währinger Straße 17, 1090 Vienna, Austria. E-mail:
leticia.gonzalez@univie.ac.at,
philipp.marquetand@univie.ac.at, christoph.dellago@univie.ac.at}}
\footnotetext{\textit{$^{b}$~Vienna Doctoral School in Physics, University of Vienna, Boltzmanngasse 5, 1090 Vienna, Austria. }}
\footnotetext{\textit{$^{c}$~Vienna Doctoral School in Chemistry, University of Vienna, Währinger Straße 42, 1090 Vienna, Austria. }}
\footnotetext{\textit{$^{d}$~Institute of Theoretical Chemistry, Faculty of Chemistry, University of Vienna, Währinger Straße 17, 1090 Vienna, Austria. }}
\footnotetext{\textit{$^{e}$~Faculty of Physics, University of Vienna, Kolingasse 14-16, 1090 Vienna, Austria. }}

%Please use \dag to cite the ESI in the main text of the article.
%If you article does not have ESI please remove the the \dag symbol from the title and the footnotetext below.
%\footnotetext{\dag~Electronic Supplementary Information (ESI) available: [details of any supplementary information available should be included here]. See DOI: 00.0000/00000000.}
%additional addresses can be cited as above using the lower-case letters, c, d, e... If all authors are from the same address, no letter is required

%\footnotetext{\ddag~Additional footnotes to the title and authors can be included \textit{e.g.}\ `Present address:' or `These authors contributed equally to this work' as above using the symbols: \ddag, \textsection, and \P. Please place the appropriate symbol next to the author's name and include a \texttt{\textbackslash footnotetext} entry in the the correct place in the list.}

%%%END OF FOOTNOTES%%%
\noindent\LARGE{\textbf{ \\ \\Supporting Information: Nonadiabatic forward flux sampling for excited-state rare events}} \\%Article title goes here instead of the text ''This is the title''
\large \\ \noindent\large{Madlen Maria Reiner,\textit{$^{ab}$} Brigitta Bachmair,\textit{$^{ac}$} Maximilian Xaver Tiefenbacher,\textit{$^{ac}$} Sebastian Mai,\textit{$^d$} Leticia González,$^\ast$\textit{$^{ad}$}} Philipp Marquetand,$^\ast$\textit{$^{ad}$} and  Christoph Dellago$^\ast$\textit{$^{ae}$} \\%Author names go here instead of ''Full name'', etc.
%%%MAIN TEXT%%%%
\tableofcontents
\FloatBarrier
\color{white}{.\\}\color{black}

\section{Implementation details}
\subsection{OPS (Open Path Sampling)}
The modified OPS\cite{Swenson2019,Swenson2019a} version used in this work includes engines for SHARC\cite{Mai2019} and SchNarc\cite{Westermayr2020}, and an optimized engine for SchNarc called \texttt{SchNarcOpt}. Furthermore, we add a path simulator called \texttt{forward\_flux}, containing different versions of FFS (forward flux sampling)\cite{Allen2005} algorithms, in particular a general one applicable to any engine, and optimized versions dedicated to the \texttt{SchNarcOpt} engine. The latter include a general one and one that can be restarted at each interface. The code is available at \url{https://github.com/MadlenReiner/openpathsampling/tree/virapid}.

\subsection{SHARC (Surface Hopping including ARbitrary Couplings)}
We include a Langevin thermostat\cite{Gronbech-Jensen2013,Gronbech-Jensen2014} in the SHARC program. The SHARC code including our modifications is available upon request and will be made publicly available in a forthcoming release.

\subsection{PySHARC (Python wrapper of SHARC)}
PySHARC is an interface used to perform SHARC simulations in a Python program. We modified the function \texttt{run\_sharc} within the class \texttt{SHARC\_INTERFACE} so that it is possible to execute single time steps instead of a whole SHARC trajectory. The modified version of the code is available on request and will also be made publicly available in a forthcoming release of SHARC.

\subsection{SchNarc (Interface between SchNet and SHARC)}
SchNarc\cite{Westermayr2020} is originally designed as an interface between SHARC and an extension to excited-state properties of the neural network potential package SchNet.\cite{Schutt2017,Schutt2018}
The module \texttt{schnarc\_md.py} is modified to incorporate the analytical models. Examples for the models discussed in the paper are provided at \url{https://github.com/MadlenReiner/SchNarc/tree/virapid/src/scripts}.
%available upon request.

\color{white}{.\\}\color{black}
\section{Analytical model systems}
\subsection{Reduced unit systems}
\label{sec:arbu}
Since our models are analytically constructed, using conventional units like femtoseconds for times or electron volts for energies is not the best solution, because it may suggest connections to real chemical systems that do not exist. Therefore, we decided to convert our results to system-specific units that are meaningful for the respective analytical models and that are internally consistent, \textit{i.e.}, comparable relative to each other within each system of units.

For the model featuring an avoided crossing (see Sec.~4.1), we choose our energy unit $\widehat{=} ~1~\epsilon$ and length unit $x_0$ such that the diabatic potential energy surfaces in $x$ are defined as
\begin{equation}
 V_{\pm}=\frac{\epsilon}{x_0^2}(x-x_0)^2
\end{equation}
\textit{i.e.}, the two diabatic PESs have minima at $x=\pm 1~x_0$ and intersect the $y$~axis at $1~\epsilon$. Masses are measured in units of the mass $m$ of the system. Times are measured in units of $\sqrt{\frac{mx_0^2}{\epsilon}}$. The unit of action in our unit system is $\sqrt{\epsilon m x_0^2}$. The temperature in a conventional unit system is accessed via the relation $\frac{k_BT}{\epsilon}=1$.

For the model featuring a conical intersection (see Sec.~4.2), we choose energy, length, time, and mass units of our unit system in the same manner, by placing our diabatic PESs minima at coordinates $(x,y)=(0.5~x_0,3.0~x_0)$ and $(x,y)=(3.0~x_0,0.5~x_0)$, and fixing one energy unit $\epsilon$ as intersect of the diabatic potentials and the $z$~axis.

The final conversion table between the unit system used by SHARC input files (\textit{i.e.}, atomic units for everything except for time that is measured in femtoseconds) and the unit systems used for the analytical model systems are shown in Table~\ref{tbl:arbu}.
\begin{table*}
\small
  \caption{\ Unit conversion table for the analytical model systems between conventional units and system-specific self-consistent units used in this work}
  \label{tbl:arbu}
  \begin{tabular*}{\textwidth}{@{\extracolsep{\fill}}lll}
    & Model system ``avoided crossing'' & Model system ``conical intersection''
    \\\hline
    energy $E$ & $[E]=1~\epsilon$ $\widehat{=}$ $0.05$~Ha & $[E]=1~\epsilon$ $\widehat{=}$ $0.078125$~Ha \\
length $L$ & $[L]=1~x_0$ $\widehat{=}$ $1$~Bohr & $[L]=1~x_0$ $\widehat{=}$ $1$~Bohr \\
mass $M$ & $[M]=1~m$ $\widehat{\approx}$ $1837.15$~m$_e$ & $[m]=1~m$ $\widehat{\approx}$ $1837.15$~m$_e$ \\
time $t$ & $[t]=1~\sqrt{\frac{mx_0^2}{\epsilon}}$ $\widehat{\approx}$ $4.6366$~fs & $[t]=1~\sqrt{\frac{mx_0^2}{\epsilon}}$ $\widehat{=}$ $3.7093$~fs\\
temperature $T$ & $[T]=1~\frac{\epsilon}{k_B}$ $\widehat{\approx}$ $15788.76$~K & $[T]=1~\frac{\epsilon}{k_B}$ $\widehat{\approx}$ $24669.93$~K \\
    \hline
    Planck's constant $\hbar$ & $\hbar\approx 0.1043379668~\sqrt{\epsilon m x_0^2}$ & $\hbar \approx 0.0834703735
~\sqrt{\epsilon m x_0^2}$\\
    \hline
  \end{tabular*}
\end{table*}

\subsection{Rare event dynamics through an avoided crossing}
The time step used in this model system is $0.25$~fs in conventional units. We use a TSH decoherence parameter $C = 2$, which in atomic units corresponds to
$0.1$~Ha.\cite{Granucci2010} Computational details for the NAFFS simulation given in Table~1 in our paper are shown in Table~\ref{tbl:1D-compdetails}. For the NAFFS simulation of half the size that is mentioned in the paper, we use $0.5$~million time steps in the flux calculation and $1000$ shots per interface, and the definition of the interfaces is not changed compared to Table~\ref{tbl:1D-compdetails}.

\begin{table}[h]
\small
  \caption{\ Computational details for the NAFFS simulation presented in Table~1 in our paper. %The flux and crossing probabilities are calculated as described in Sec.~\ref{sec:FFS}
  }
  \label{tbl:1D-compdetails}
  \begin{tabular*}{1.0\textwidth}{@{\extracolsep{\fill}}llll}
    \hline
    flux sim. & time steps & crossing events & flux ($10^{-3}$) \\
    \hline
    & $1$~million & $2137$ & $(77.1\pm 1.7)$ \\
    \hline
    NAFFS sim. & interfaces in $x$ & shots per interface & crossing prob. \\
    \hline
    & $-0.5\rightarrow -0.2$ & $2000$ & $(24.6\pm1.0)\%$ \\
    & $-0.2\rightarrow 0.0$ & $2000$ & $(50.9\pm1.2)\%$ \\
    & $0.0\rightarrow 0.5$ & $2000$ & $(90.25\pm0.67)\%$ \\
    \hline
  \end{tabular*}
\end{table}

\begin{figure*}
    \centering
    \includegraphics[height=5.0cm]{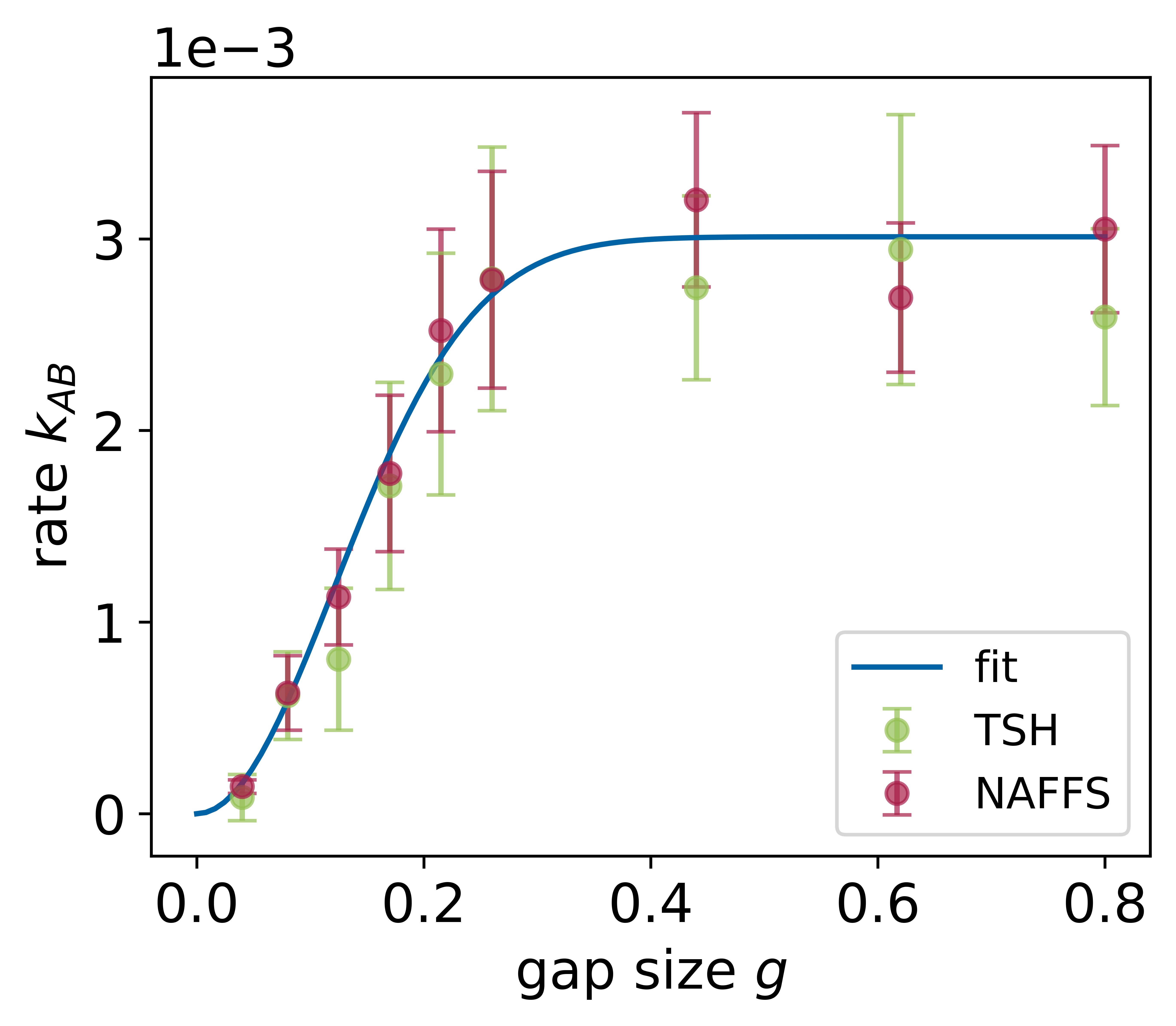}
    \caption{Brute-force TSH rates as a function of the gap size $g$ compared to Fig.~5b in our paper.}
 \label{fig:1D-gapsize-TSH}
\end{figure*}

Brute-force TSH results accompanying Fig.~5b in our paper are shown in green in Fig.~\ref{fig:1D-gapsize-TSH}. They agree very well with our NAFFS results.

\subsection{Rare event dynamics through a conical intersection}
The time step used in this model system is $0.5$~fs in conventional units. We use a decoherence parameter $C = 1.28$, which in atomic units corresponds to $0.1$~Ha.\cite{Granucci2010}

Computational details for the NAFFS simulation given in Table~3 in our paper are shown in Table~\ref{tbl:2D-compdetails}
\begin{table*}[h]
\small
  \caption{\ Computational details for the NAFFS simulation presented in Table~3 in the main paper.}
  \label{tbl:2D-compdetails}
  \begin{tabular*}{1.0\textwidth}{@{\extracolsep{\fill}}llll}
    \hline
    flux sim. & time steps & crossing events & flux ($10^{-3}$) \\
    \hline
    & $1$~million & $6527$ & $(100.1 \pm 2.8)$ \\
    \hline
    NAFFS sim. & interfaces in $(x-y)$ & shots per interface & crossing prob. \\
    \hline
    & $2.5\rightarrow 0.0$ & $2000$ & $(29.9\pm 1.1)\%$ \\
    & $0.0\rightarrow -1.5$ & $2000$ & $(37.8\pm 1.1)\%$ \\
    & $-1.5\rightarrow -2.5$ & $2000$ & $(51.3\pm 1.2)\%$ \\
    \hline
  \end{tabular*}
\end{table*}

\begin{figure*}
    \centering
    \includegraphics[height=10.0cm]{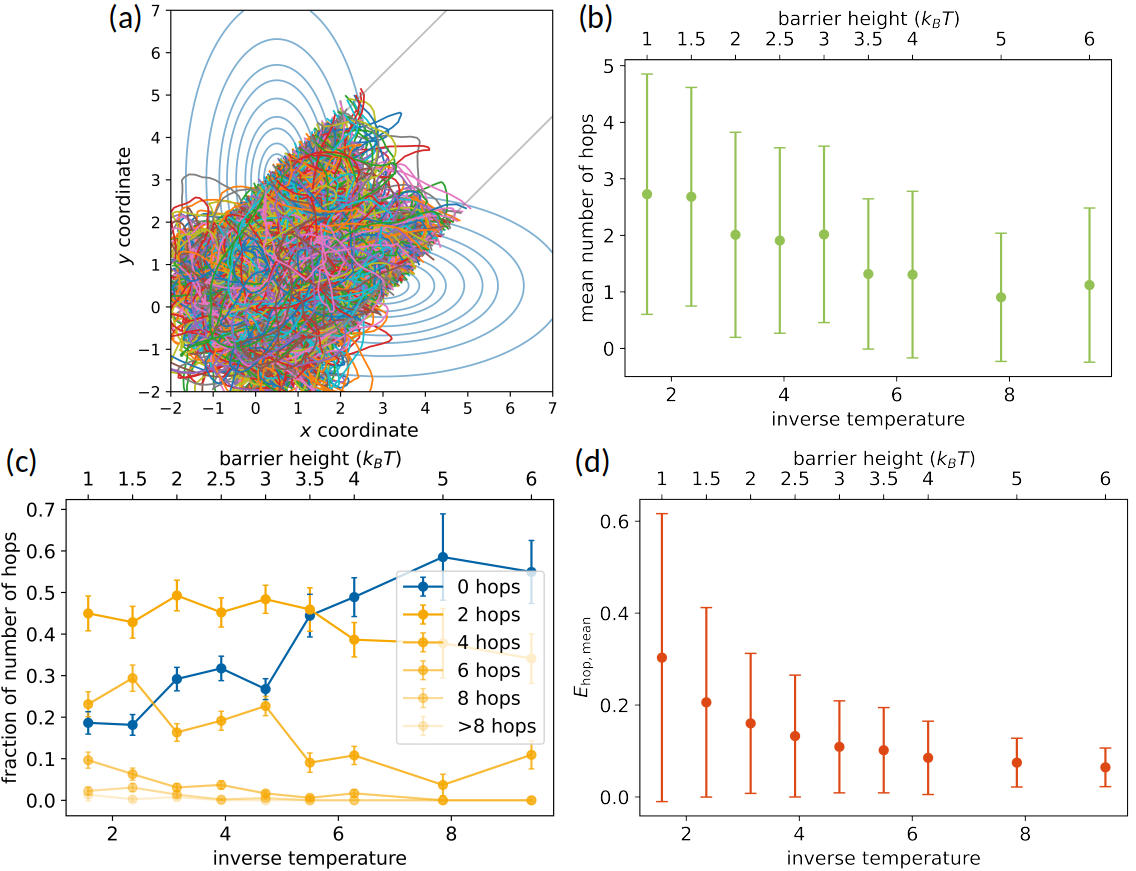}
    \caption{(a) 1857 transition paths obtained in the TSH simulation of the two-dimensional analytical model. Adiabatic ground state contour lines are shown in blue, and the boundaries of the initial and final regions are shown in gray. (b) The mean number of hops per NAFFS transition path as a function of the inverse temperature, and, alternatively, as a function of the barrier height. The error bars show the obtained standard deviation and do not indicate the accuracy of the obtained results. (c) The ratio of NAFFS transition paths showing a certain number of hops, where the $x$~axes are the same as in (b). The values of the error bars $\Delta f$ are calculated with respect to the number of paths $N$ showing the specific number of hops and their ratio $f$ with respect to the total number of transition paths, \textit{i.e.}, $\Delta f = f\cdot \frac{\sqrt{N}}{N} $. (d) The average potential energy of a hop from the ground state to the excited state in a TSH simulation, where the $x$~axes equal the ones in (b). The error bars indicate the obtained standard deviation.}
 \label{fig:2D}
\end{figure*}

For qualitative comparison of the reactive paths of the TSH and NAFFS comparisons presented in our work, we show here (see Fig.~{\ref{fig:2D}}a) the transition paths obtained in the TSH simulation consisting of $5$~million time steps that is described in the paper (see Sec.~4.2). They qualitatively agree with the ones sampled in the NAFFS simulation, which are shown in Fig.~6d in our paper. Note that almost twice as many reactive paths are plotted in case of the TSH simulation ($1857$, see Fig.~\ref{fig:2D}a) than by the NAFFS simulation ($1025$, see Fig.~6d in the paper), resulting in the TSH paths being denser. Note that the initial and final regions $A$ and $B$ are defined not only geometrically as plotted in Fig.~\ref{fig:2D}a, but also in terms of the PES, namely on the ground state. Accordingly, paths that seem to sojourn in $A$ or $B$ in the plot are located on the excited state -- and, hence, neither in $A$ nor in $B$. %The two calculations also agree in the average number of completed hops in adiabatic representation of transition paths.

An evaluation of the average number of hops in transition paths as a function of the inverse temperature shows a decreasing behavior (see Fig.~\ref{fig:2D}b). In accordance to that we find that the fraction of paths which undergo the transition purely on the ground state with respect to the total number of reactive paths increases with decreasing temperature, whereas the ratio of transition paths that show hops to the excited state decreases with decreasing temperature (see Fig.~\ref{fig:2D}c). Fig.~\ref{fig:2D} also shows the ratio of reactive paths that have more than eight hops in a transition. Only in the three calculations with the highest temperatures this fraction is greater than zero. Where such hops occur, their number is very small (only $29$ hops in all calculations combined), both of which militate against examining these numbers in more detail statistically as they are not significant, so we limit ourselves to the paths that show eight hops or fewer.

Fig.~\ref{fig:2D}d shows the mean energy of a hop from the ground state to the excited state as a function of the inverse temperature. It is obtained as the potential energy difference at the geometries where the hops occur. As expected, it decreases with decreasing temperature, since energy must be spent for hops, and the system has less energy available at a lower temperature. At higher temperature, both hops that require a lot of energy and those that cost little energy can occur, while at low temperature, only those hops that require little energy are possible. Accordingly, the standard deviation of the mean energy decreases with decreasing temperature.

%%%END OF MAIN TEXT%%%


\balance

\FloatBarrier
%%%REFERENCES%%%
\bibliography{library} %You need to replace "rsc" on this line with the name of your .bib file
\bibliographystyle{rsc} %the RSC's .bst file